\documentclass[12pt,preprint]{aastex}
\usepackage{natbib,graphicx,latexsym}
\newcommand{\lsim}{\
\raise-2.truept\hbox{\rlap{\hbox{$\sim$}}\raise5.truept\hbox{$<$}\ }}
\newcommand{\gsim}{\
\raise-2.truept\hbox{\rlap{\hbox{$\sim$}}\raise5.truept\hbox{$>$}\ }}

\begin{document}
\shorttitle{...}
\title{Detection of Surface Brightness Fluctuations in Elliptical
Galaxies imaged with the Advanced Camera for Surveys. B- and I-band
measurements\altaffilmark{1}}

\author{Michele Cantiello\altaffilmark{2,3}, Gabriella Raimondo\altaffilmark{3},
John P.~Blakeslee\altaffilmark{2}, Enzo Brocato\altaffilmark{3},
Massimo Capaccioli\altaffilmark{4,5}}
\altaffiltext{1}{Based on observations made with the NASA/ESA Hubble Space
Telescope, which is operated by the Association of Universities for
Research in Astronomy, Inc., under NASA contract NAS 5-26555. These
observations are associated with programs \#9427.}
\altaffiltext{2}{Department of Physics and Astronomy, Washington State University,
Pullman, WA 99164.}
\altaffiltext{3}{INAF--Osservatorio Astronomico di
Teramo, Via M. Maggini, I-64100 Teramo, Italy}
\altaffiltext{4}{Dipartimento di Scienze Fisiche, Universit\`a Federico II di
Napoli, Complesso Monte S. Angelo, via Cintia, 80126, Napoli, Italy}
\altaffiltext{5}{INAF--Osservatorio Astronomico di Capodimonte, via
Moiariello 16, 80131 Napoli, Italy}

\begin{abstract}

Taking advantage of the exceptional capabilities of ACS on board of
HST, we derive Surface Brightness Fluctuation (SBF) measurements in
the B and I bands from images of six elliptical galaxies with
$1500~\leq~cz \leq 3500$. Given the low S/N ratio of the SBF signal
in the blue band images, the reliability of the measurements is
verified both with numerical simulations and experimental data
tests.

This paper presents the first published B- and I-band SBF
measurements for distant ($\geq$ 20 Mpc) galaxies, essential for the
 comparisons of the models to observations of
normal ellipticals. By comparing I-band data with our new Simple
Stellar Population (SSP) models we find an excellent agreement and
we confirm that I-band SBF magnitudes are mainly sensitive to the
metallicity of the dominant stellar component in the galaxy, and are
not strongly affected by the contribution of possible secondary
stellar components. As a consequence I-band fluctuations magnitudes
are ideal for distance studies. On the other hand, we show that
standard SSP models do not reproduce the B-band SBF magnitudes of
red ( $(B-I)_0 \gsim 2.1$ ) galaxies in our sample. We explore the
capability of two non--canonical models in properly reproducing the
high sensitivity of B SBF to the presence of even small fractions of
bright, hot stars (metal poor stars, hot evolved stars, etc.). The
disagreement is solved both by taking into account hot (Post--AGB)
stars in SSP models and/or by adopting Composite Stellar Population
models. Finally, we suggest a limit value of the S/N for the B-band
SBF signal required to carry out a detailed study of stellar
population properties based on this technique.

\end{abstract}

\keywords{galaxies: elliptical and lenticular, cD -- galaxies: photometry
-- galaxies:  evolution -- galaxies: stellar component}

\section{Introduction}
The first applications of the Surface Brightness Fluctuations (SBF)
technique were mainly devoted to refine the method itself, and to
gauge distances of elliptical galaxies and bulges of spirals up to
$\sim$ 20 Mpc \citep[e.g.,][]{ts88,tonry89,tonry90,
tonry91,jacoby92}. A deeper understanding of the SBF technique,
together with technological improvements of telescopes, made it
possible to extend the method to distances as far as $\sim 100$ Mpc
\citep{jensen01} with uncertainties typically lower than 10\%, and to
measure luminosity fluctuations for a wider family of astronomical
objects ranging from Galactic and Magellanic Clouds globular clusters
\citep{at94,gonzalez04,raimondo05}, to dwarf ellipticals
\citep{jerjen98,jerjen00,mieske03}, in addition to the usual targets
(elliptical galaxies and bulges spirals).

In their seminal paper \citeauthor{ts88} suggested that it was also
possible to take advantage of SBF to analyze the metallicity and age
of the stellar system in the galaxy. Such connection was revealed
through the correlation between the SBF amplitudes and the integrated
color of the galaxy.  This behavior, effectively detected only after a
substantial number of measurements was available \citep{tonry90}, on
one hand showed that to obtain reliable SBF-based distances SBF versus
integrated color relations must be used to ``standardize'' the SBF
absolute magnitude. On the other hand, it also emphasized the possible
use of the SBF as a tracer of stellar population properties.

The link between SBF and stellar populations is easily understood
considering that, given the image of a galaxy, the SBF signal rises
from the statistical fluctuation between adjacent regions, due to the
finite number of stars per resolution element. In particular the SBF
is defined as the spatial fluctuation of the galaxy surface
brightness, normalized to the surface brightness itself. As
demonstrated by \citeauthor{ts88}, by definition the SBF amplitude
corresponds to the ratio of the second to the first moments of the
luminosity function of the underlying stellar system. This means that,
for a typical stellar population of an early-type galaxy composed
mainly by old and metal rich stars ($t\gsim 5$ Gyr,
[Fe/H]$\gsim$-0.3), the SBF nearly corresponds to the average
magnitude of the red giant branch (RGB) stars in the population. In
this view -- keeping in mind that stellar systems with different
metallicities and ages have different RGBs, thus different SBF -- the
connection between SBF and stellar populations is clear.

In order to use SBF measurements to analyze the age and chemical
composition of stellar systems, when individual stars cannot be
resolved, there are two facts to be considered. First, on the
observational side, in addition to the apparent SBF magnitude an
estimation of the galaxy distance is needed so that the absolute SBF
magnitude can be derived; otherwise, SBF color or SBF gradient data,
which are distance-independent, are needed. Second, theoretical values
for the absolute SBF magnitudes must be available.

The theoretical study of fluctuations amplitudes has been widely
explored in the last decade, typically using Simple Stellar Population
models (SSP). Since \citet{tonry90}, many other authors
\citep[e.g.,][]{buzzoni93,worthey93a,bva01,liu00,cantiello03,mouhcine05,raimondo05,marin06}
explored a wide range of metallicities ($0.0001 \lsim Z \lsim 0.05$),
ages ($20~Myr \lsim t \lsim 18~Gyr$), in various wavelength
intervals. These models showed a general agreement, though they are
based on quite different input stellar physics. A general conclusion
drawn from all models is that SBF magnitudes at shorter wavelengths
intervals, e.g. in the B-band, show a stronger sensitivity to the
chemical composition and age with respect to other bands, thus they
must be preferred to address stellar population studies. This is
particularly interesting in objects like elliptical galaxies, where
dust pollution (which strongly affects bluer bands) is negligible or
recognizable as irregular spots in the otherwise regular galaxy
profile.

Up to now, the few examples on the use of SBF to probe the unresolved
stellar content of galaxies have been based mainly on the comparison
with models of the absolute fluctuation magnitudes $\bar{M}_{\lambda}$
in a certain wavelength interval [$\lambda$, $\lambda + \Delta
\lambda$], thus relying on some assumption of the galaxy
distance. There are fewer studies based on SBF color data, due to the
general lack of multi-wavelength SBF data for the same
galaxy. One recent application to galaxies is the work by
\citet{jensen03} which couples ground-based I-band SBF with near-IR
(F160W) HST data. Finally, only a few works presented SBF radial
gradients within galaxies
\citep{tonry91,sodemann95,tonry01,cantiello05}.

The aim of this paper is to present a set of B- and I-band SBF
measurements for a selected sample of galaxies imaged with the ACS
camera on board of HST. In \citet[C05 hereafter]{cantiello05} we
succeeded in revealing $\bar{M}_I$ radial ($10 \lsim r (arcsec) \lsim
40 $) gradients within 7 early-type galaxies.  For six of these
galaxies F435W ($\sim$ Johnson B-band) images are also
available. Taking advantage of the capabilities of the ACS camera, in
this paper we present the SBF analysis for the B images, whose SBF
signal is expected to be substantially fainter than in the I-band
images. To produce a homogeneous set of measurements for both bands,
we repeat the SBF analysis also for I-band data adopting the same
constraints (masks, regions for the SBF measurements, etc.) as for the
B images.

The paper is organized as follows. Section 2 describes the selected
sample of objects, the procedures adopted to derive the surface
photometry, the photometry of point-like and extended sources, and the
SBF magnitudes. In section 3 we introduce some tests performed to
check the quality of our SBF measurements on B-band frames. The
analysis of some observational properties of the galaxies sampled
(plus two more objects whose data are taken from the literature), and
the comparison of data with models are presented in section 4.  We
finish this paper with the conclusions, in section 5.

\section{Data reduction and analysis}

\subsection{Observations}

Five of the galaxies considered here are the same presented in
C05 (NGC\,1407, NGC\,3258, NGC\,3268, NGC\,5322 and NGC\,5557),
while NGC\,404 and NGC\,1344 were excluded because no F435W images
are available. The raw data are ACS F435W, and F814W exposures
drawn from the HST archive. All data are taken with ACS in its
Wide Field Channel mode. The observations are deep exposures
associated with the proposal ID \#9427, which was designed to
investigate the Globular Cluster System for a sample of 13 giant
ellipticals in the redshift regime $1500\leq cz \leq 5000$. Among
these galaxies we selected six objects less polluted by dust
patches and with exposure times long enough to allow SBF
measurements. Table \ref{tab_data} summarizes some properties of
the galaxies sampled together with the exposure times in each
filter.

As in C05, the image processing (including cosmic-ray rejection,
alignment, and final image combination) is performed with the
``Apsis'' data reduction software \citep{blake03}.  The ACS
photometric zero points and extinction ratios are from
\citet[hereafter S05]{sirianni05}, along with a correction for
Galactic absorption from \citet{sfd98}.  Because of the small quantity
of dust in normal elliptical galaxies, no internal extinction
correction has been considered here. The dusty patches present in the
case of NGC\,4696 can be easily recognized in the B-band images and
have been masked out. For this galaxy the percentage of the image area
masked for dust is $<$2\% of the whole ACS field of view (or $<8$\%
the final area selected for SBF measurement).

To transform the ACS photometry to the standard UBVRI photometric
system, we use the equations from S05. For the B-, and I-band data the
average difference between the ACS magnitudes and the ones obtained
after applying the S05 equations does not exceeds 0.03 mag.  In C05 we
presented some consistency checks of the (F435W, F814W)-to-(B,I)
transformations, by comparing for each galaxy the transformed
magnitudes and colors with available measurements taken in the
standard photometric system. The results of the comparisons supported
the reliability of S05 equations for the selected filters (see C05 for
more details).

\subsection{Data analysis}
The procedure adopted to derive SBF measurements from the B-band
frames is the same described in C05, with minor changes which will be
discussed below together with a brief summary of the whole procedure.

\subsubsection{Galaxy modeling and sky subtraction}

Galaxy light profile modeling and subtraction, sky subtraction, and
sources masking proceeded in an iterative way.  Briefly, a
provisional sky value for the whole image is assumed to be equal to
the median pixel value in the corner with the lowest number of
counts. After this sky value has been subtracted from the original
image, we started an iterative procedure where: (i) the fit of the
median galaxy profile is obtained using the IRAF/STSDAS task
ELLIPSE; (ii) all the foreground stars, background galaxies, and the
galaxy's globular clusters ({\it external sources} hereafter)
detected in the sky+galaxy subtracted frame are masked out; (iii) a
new sky value is obtained by fitting a $r^{1/4}$ law to the surface
brightness profile of the galaxy.

The procedure is then repeated, each time the ellipse fitting is
performed using the updated mask of external sources, until
convergence. As shown in C05, the uncertainty of assuming a $r^{1/4}$
profile instead of a $r^{1/n}$ or a Nuker profile, does not alter
significantly the final value of integrated and SBF magnitudes, at
least in the regions where the galaxy's signal is higher than the
background. The galaxy counts in the regions where the SBF color
measurements were made are on average a factor $\sim$7 ($\sim$17)
higher than the sky level in the B (I) images.

\subsubsection{Photometry of point-like and extended sources}

Once the sky and the galaxy model were subtracted from the original
image, we derived the photometry of the external sources.  The
construction of the photometric catalogue of point-like and extended
sources is critical for the estimation of the Luminosity Function
(LF). As is shown in C05, by fitting the LF of external sources
(i.e. of globular clusters and background galaxies) we can infer, by
extrapolation, its faint end, which must be determined to have
reliable SBF measurements. This is due to the fact that the
fluctuations measured from the sky+galaxy subtracted frame also
include a contribution arising from the undetected external sources
left in the image. A reliable LF model allows one to properly estimate
and subtract such residual extra-fluctuation (usually indicated as
$P_r$) from the total fluctuations signal, $P_0$.

To obtain the photometry of external sources we used the software
SExtractor \citep{bertin96}, as it gives good photometry of both
point-like and extended objects.  It also gives as output a smoothed
background map which is essential to remove the large scale residuals
from the galaxy subtracted frame. Moreover, it accepts user specified
weight images, and this is of great importance for our measurements,
as, by specifying the error map, the photometric uncertainty can be
estimated taking into account the contribution to the noise due to the
subtracted galaxy. We have modified the SExtractor input weight images
by adding to the RMS image the galaxy model times a constant factor
$\sim0.5$ for B-, and $\sim 1$ for I-band images. The constant factors
have been chosen, after several checks, so that the surface brightness
fluctuations are recognized as noise \citep{tonry90}, lower
coefficient values would result in the detection and masking of
fluctuations as external sources; on the contrary, higher values would
result in many real external sources being undetected and therefore
affect the SBF measurement.

We ran SExtractor on the residual images (i.e., galaxy+sky+large scale
residuals subtracted images) independently for each band.  The best
parameters for source detection have been chosen by using numerical
simulations. In particular, we have simulated images with known input
LF, then the simulated images have been analyzed in the same way of
real frames to study the LF of the sources detected. Finally the input
LF have been compared with the ones derived from our standard analysis
of the frame.

\begin{figure}
\begin{center}
\epsscale{1.}
\plottwo{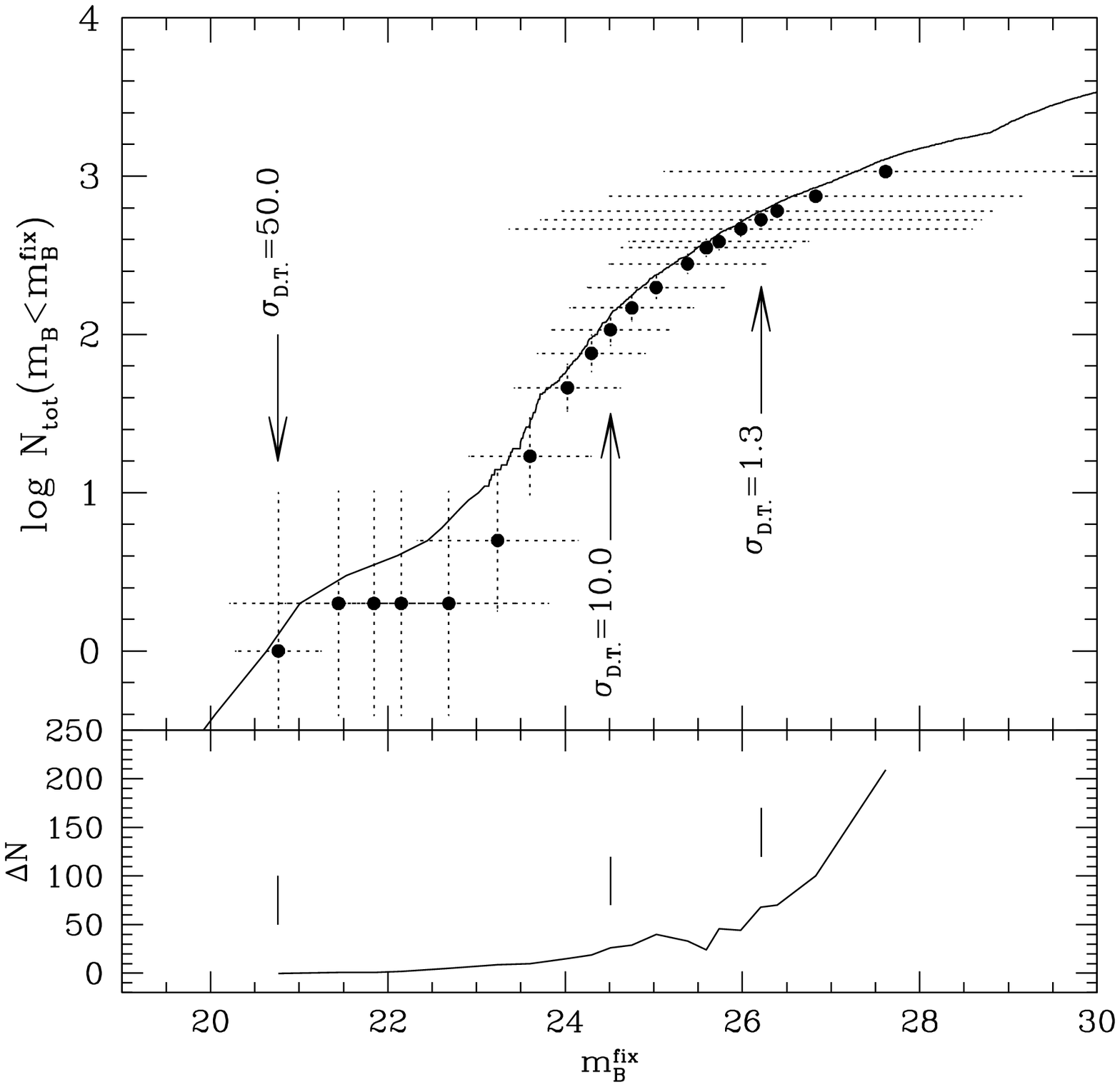}{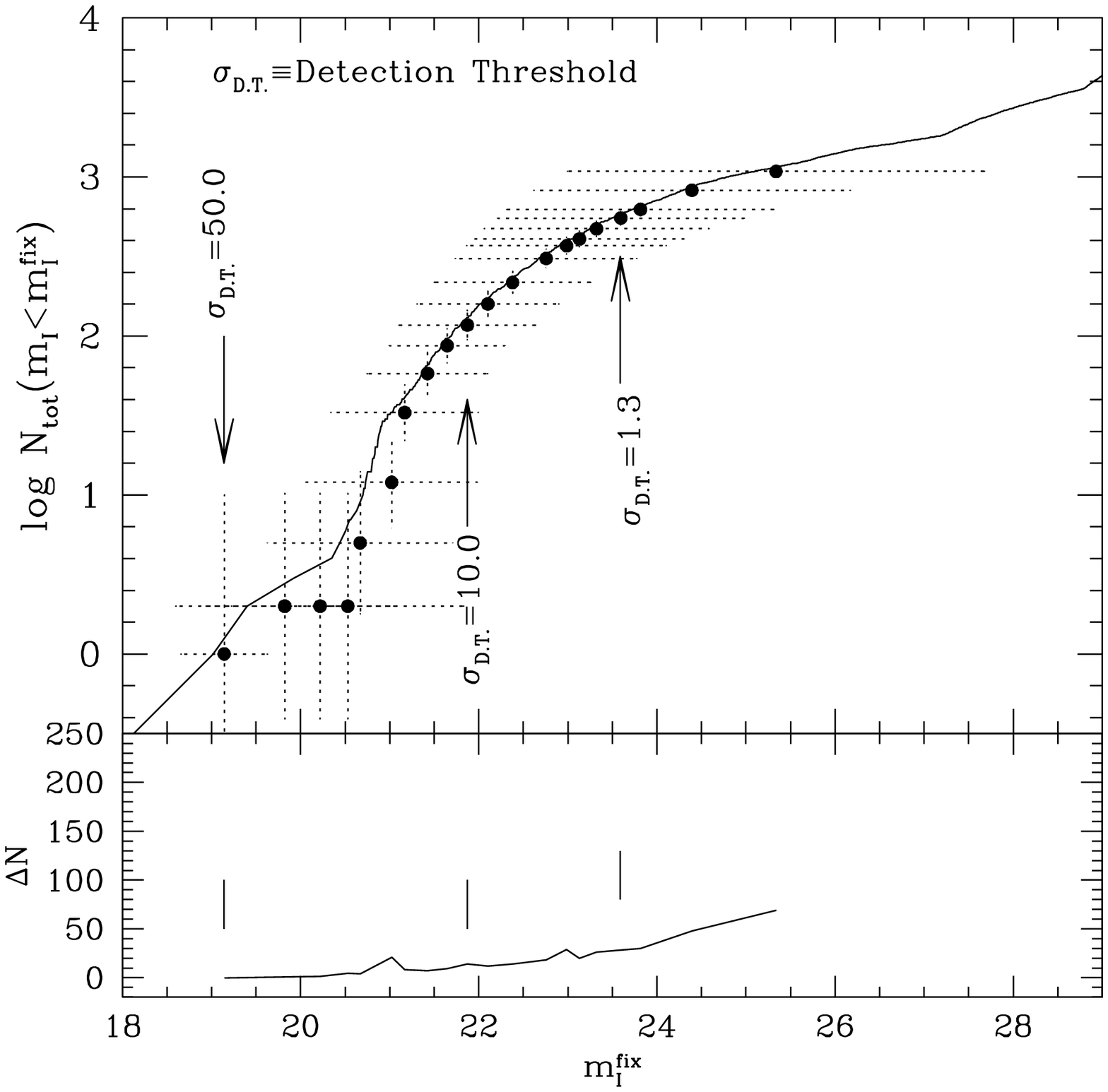}
\figcaption{{\it Upper panels} - The number of sources detected
(globular clusters+galaxies) brighter of the magnitude $m^{fix}$, versus the
magnitude $m^{fix}$ itself. We defined the latter quantity as the
average magnitude of the sources detected for the specific
$\sigma_{D.T.}$ adopted. Left (right) panels exhibits the results of
this test for B-band (I-band) simulations. Full dots refer to the
total number of objects detected on the frame simulated, the solid line
shows the input total integrated luminosity function adopted for the
simulations. {\it Lower panels} - The differences between the number
of detected objects brighter than $m^{fix}$, and the real number of
objects adopted for the simulation: $\Delta N = |N_{detected} -
N_{input}|$.
\label{sextest}}
\end{center}
\end{figure}

As an example of this procedure, in Figure \ref{sextest} we show the
derived best--fit of the LF of external sources obtained to determine
the optimal {\it Detection Threshold} (DETECT\_THRESH, $\sigma_{D.T.}$
hereafter) SExtractor parameter\footnote{ We have used a fixed
DETECT\_MINAREA$\sim$5, corresponding to the PSF area of our ACS
images. The SExtractor detection S/N limit is $\sigma_{D.T.}  \times
\sqrt(DETECT\_MINAREA)$. For this test we have used the images
simulated according to prescriptions presented in section 3. This
allowed us to compare the observed integrated luminosity function with
the input one.}.

In detail, we changed $\sigma_{D.T.}$ within the interval 0.5-50,
obtaining different photometric catalogs, each one characterized by
the number of sources detected in both frames, and by the average
magnitude of the sources detected $m^{fix}$. Lowering the
$\sigma_{D.T.}$ one has correspondingly ($i$) an increasing number of
sources detected, ($ii$) a fainter average magnitude detected
$m^{fix}$, and ($iii$) an increasing number of spurious detections. In
the Figure \ref{sextest} we plot with full dots the total observed
number (N$_{detected}$) of detected sources brighter than $m^{fix}$
versus $m^{fix}$ for B (upper left panel), and I band (upper right
panel). The solid line in these panels shows the input total number of
sources (N$_{input}$). The lower panels show the residuals, i.e. :
$\Delta N =|N_{detected} - N_{input}|$.  We have chosen as final
detection parameter $\sigma_{D.T.}=1.3$ as for this value $\Delta N$
remains below 50 for both I-, and B-band frames, i.e., below $\sim 5
\% $ of the total number of sources detected (for high number of
detections).

The final photometric catalogs of external sources from B- and
I-band images were matched using a radius $0.1\arcsec$.  On
average, the number of objects clearly detected in the I-band
frame and undetected in the B-band one, e.g. red galaxies, is
$\lsim 30$. We have verified that including the photometry of
these missing objects does not significantly alter the LF model
and therefore the correction to the SBF signal for undetected
sources is not significantly changed.

Once the catalogue of matched sources was constructed, we applied the
aperture correction as in C05. That is, for point-like sources we
derived the aperture correction by making a growth-curve analysis over
few bright, well isolated point sources in the frame
\citep{stetson90}. The aperture correction was obtained by summing the
contribution evaluated from the growth-curve analysis, with aperture
diameters 6-20 pixels, and the contribution from 20 pixels to
``infinite'' diameter reported by S05. For extended sources, instead,
we used the aperture correction following the prescriptions by
\citet{benitez04}.

A model LF was then derived from the final photometric catalogue by assuming
the total number density to be the sum of a gaussian-shaped Globular
Cluster LF \citep[GCLF,][]{harris91}:
\begin{equation}
n_{GC}(m) =\frac{N_{0,GC}} {\sqrt{2 \pi \sigma^2}}~~e^{- \frac {(m - m_{\lambda,GC})^2}{2 \sigma ^2}}
\end{equation}
and a power-law LF \citep{tyson88} for the background
galaxies:
\begin{equation}
n_{gxy}(m) =N_{0,gxy} 10 ^ {\gamma m}
\end{equation}
where $N_{0,GC}$ \citep[$N_{0,gxy}$,][]{blake95} is the globular
cluster (galaxy) surface density, and $m_{\lambda,GC}$ is the turnover
magnitude of the GCLF at the galaxy distance. In expression (2) we
used the $\gamma$ values obtained by \citet{benitez04}. For the GCLF
we assumed the turnover magnitude and the width of the gaussian
function from \citet{harris01}. To fit the total LF we used the
software developed for the SBF distance survey; we refer the reader to
\citet{tonry90} and \citet{jacoby92} for a detailed description of the
procedure. Briefly: a distance modulus ($\mu_{0}$) for the galaxy is
adopted in order to derive a first estimation of
$m_{\lambda,GC}=\mu_0+M_{\lambda,GC}$, then an iterative fitting process is
started with the number density of galaxies and GC, and the galaxy
distance allowed to vary until the best values of $N_{0,GC}$,
$N_{0,gxy}$ and $m_{\lambda,GC}$ are found via a maximum likelihood method.

Figure \ref{lumfunc} exhibits the observed B- and I-band LF for the
whole sample of galaxies, and their best fit curves. In this figure we
report the LF over the entire area analyzed, although the SBF analysis
is conducted independently in annuli \citep{tonry90}. 

The source catalogue is then used to mask all sources brighter than a
(radially dependent) completeness limit.  The residual fluctuation
amplitude due to the remaining undetected faint sources, $P_r$, is
then evaluated as described in C05.  

As shown in Table \ref{tab_pi} I-band SBF magnitudes are not strongly
affected by the $P_r$ correction, which is typically a factor 20
smaller than the stellar fluctuations amplitude, $P_f$. The case of
B-band SBF is quite different. In this case, the amplitude of the
stellar SBF competes with the residual variance from undetected
sources.  For B-band measurements, in fact, we find an average $P_r
\sim 0.22 P_f$.  However, even in the case of B-band data, the
accuracy of photometric catalogue allowed to obtain a reliable LF
model.

In order to estimate the uncertainty associated to $P_r$ we have
carried out several numerical experiments by using the tool described
in forthcoming section 3. In particular, we have estimated $P_r$ from
simulations of ACS images having similar properties to our real images
and compared it with the known input value. We obtained that a
realistic estimation of the $P_r$ uncertainty is $\sim$ 20\% the value
of $P_r$. In addition we have also estimated that changes up to 50\%
of the original value in the LF fitting parameters ($\gamma$,
$M_{\lambda,GC}$, etc.) affect the final $P_r$ by less than 15\%. In
conclusion we adopt 20\% of $P_r$ as an estimate of its uncertainty.

\subsubsection{SBF measurements}

The SBF analysis executed for B-band frames is similar to the one
described in C05, except for the fact that here we derive one single
SBF value per galaxy (instead of different SBF measurements in
several concentric annuli as in C05) since B-band images do not
achieve a sufficiently high S/N to make a detailed study of the
fluctuation amplitude as a function of radius.

For sake of homogeneity we also re-analyzed I-band images by
adopting the same annuli shape as from the B frames. The annulus
was created using a mask that matches the ellipticity and position
angle of a given region.
All frames were analyzed in the same way.
After subtracting the sky, the galaxy model, and the large scale
residuals, we derived (i) the photometry of point-like and
extended sources in the frame, masking (ii) all sources above a
fixed S/N ratio ($\sim 3.0$).  Afterwards, the residual frame
divided by the square root of the galaxy model was Fourier
transformed (iii), and the azimuthal average of the power-spectrum
$P(k)$ evaluated.  Then we derived (iv) the constants $P_0$ and
$P_1$ in the equation:

\begin{equation}
P(k)= P_0 \cdot E (k) + P_1 \,,
\label{eqpk}
\end{equation}
adopting a robust linear least squares method \citep{press92}. In
the latter equation $E(k)$ is the azimuthal average of the
convolution between the PSF and the mask power spectra. The PSF
used were the template PSF from the ACS IDT, constructed from
bright standard star observations. Finally, the SBF amplitude in
magnitude was obtained as:

\begin{equation}
\bar{m}=-2.5~log(P_0-P_r) + mag_{zero} - A_{\lambda}.
\label{eqsbf}
\end{equation}

Here $P_r$ is the residual variance due to unmasked sources in the
frame, which has been estimated from the LF of the external sources
detected in the frame (section 2.2.2, and C05). The constant
$mag_{zero}$ is the zeropoint magnitude from S05, while $A_{\lambda}$
is the extinction correction.

To test the stability of these measurements, we performed several
tests. As an example, we have chosen few well isolated point sources
available in the frames, repeating the fitting operations for
eq. (\ref{eqpk}) using these new objects as PSF references.  The
resulting SBF magnitudes agree with the previous measurements within
uncertainties.  The final $P_0$, $P_1$, $P_r$ and $P_f$ values are
reported in Table \ref{tab_pi}, while the SBF magnitudes and $(B-I)_0$
color are reported in Table \ref{tab_bi}, together with other
properties of the galaxies. The SBF error estimates includes: ($i$) the
error in the determination of $P_0$, ($ii$) a $\sim$5\% error from the
PSF (C05), and ($iii$) the uncertainty of $P_r$, all summed in
quadrature. As shown in C05, the effect on SBF arising from sky
uncertainty is negligible.

The distance moduli reported in Table \ref{tab_bi} are derived
averaging the group distance moduli estimations of FP and IRAS
velocity maps distribution (Table 6 data in C05). For NGC\,5557, since
no group distance is known, we adopted the weighted average distance
estimations for the galaxy itself. M\,32 distance is derived as
weighted averages of group distances from the \citet{ferrarese00}
database, excluding SBF based distances. NGC\,5128 distance comes from
the recent \citet{ferrarese07}, based on Cepheids variables.

\section{Checks on SBF measurements reliability}

As discussed in \citet{jensen96}, and \citet{mei01}, in order to
obtain reliable SBF measurements in the IR, the S/N$\equiv (P_0 -
P_r)/P_1= P_f/P_1$ must be $\gsim$5. The conclusions of these authors
agree with \citet{blake99} who, from an observational point of view,
provided the equation to predict SBF S/N ratio, suggesting to keep
this value above 5-10.  As shown in Table \ref{tab_pi}, the I-band S/N
measured from our data is $\gsim 20$, while for B-band images it can
be as low as $\sim 2.6$, and it is $\sim 5.7 $ in the best case.  In
this section we present various consistency tests useful to verify the
reliability of our SBF measurements procedure in such low S/N regimes.

As a first check, we developed a procedure capable of simulating
realistic CCD images of galaxies with known input SBF magnitudes
derived using SSP models, then we measured the SBF of the
simulated image to verify the matching with the input SBF value.
In order to properly simulate the properties of our real images,
the simulations also included external sources, and the
instrumental noise of the camera. A detailed description of the
simulations is reported in Appendix A.

\begin{figure}
\epsscale{1} \plottwo{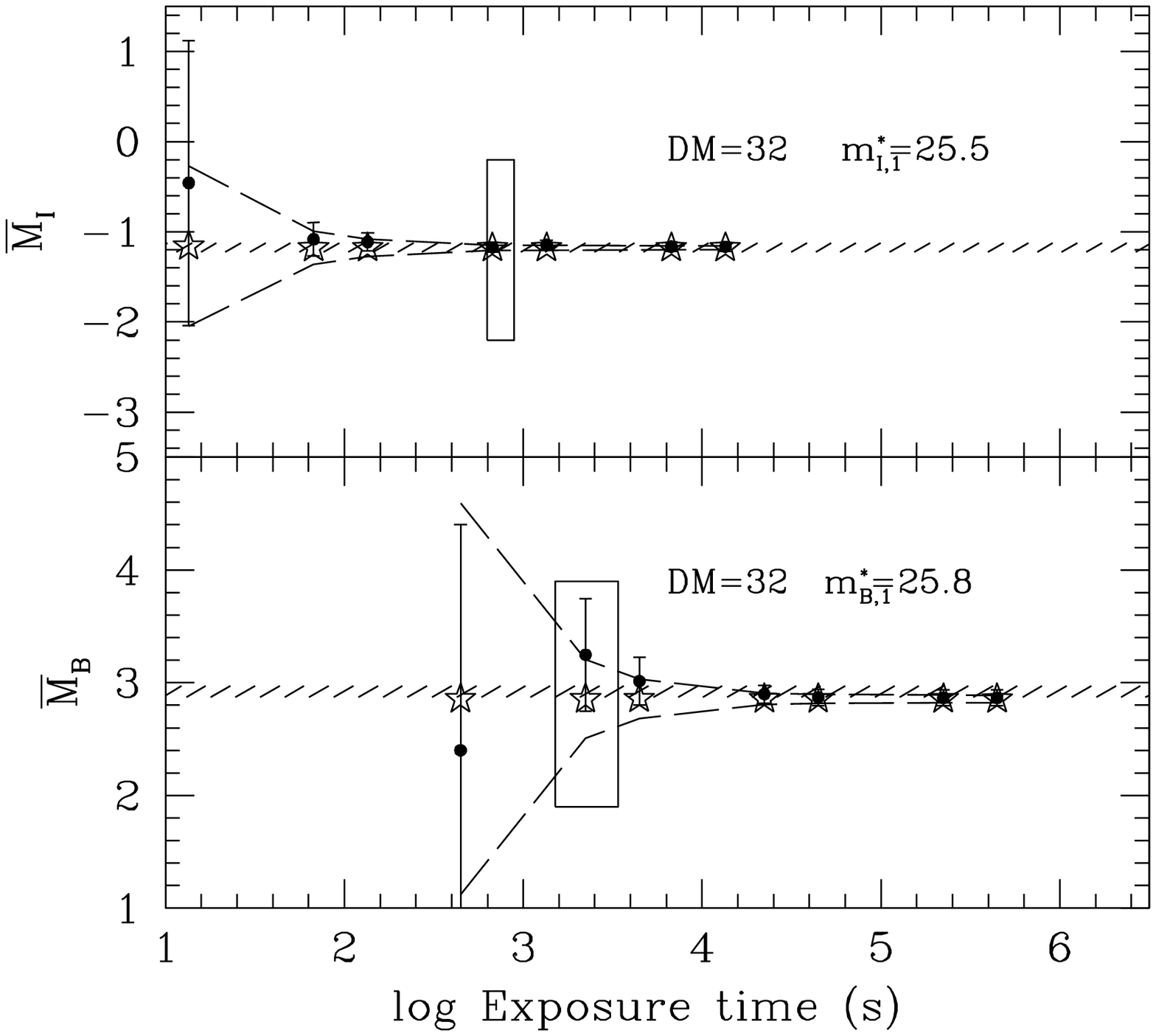}{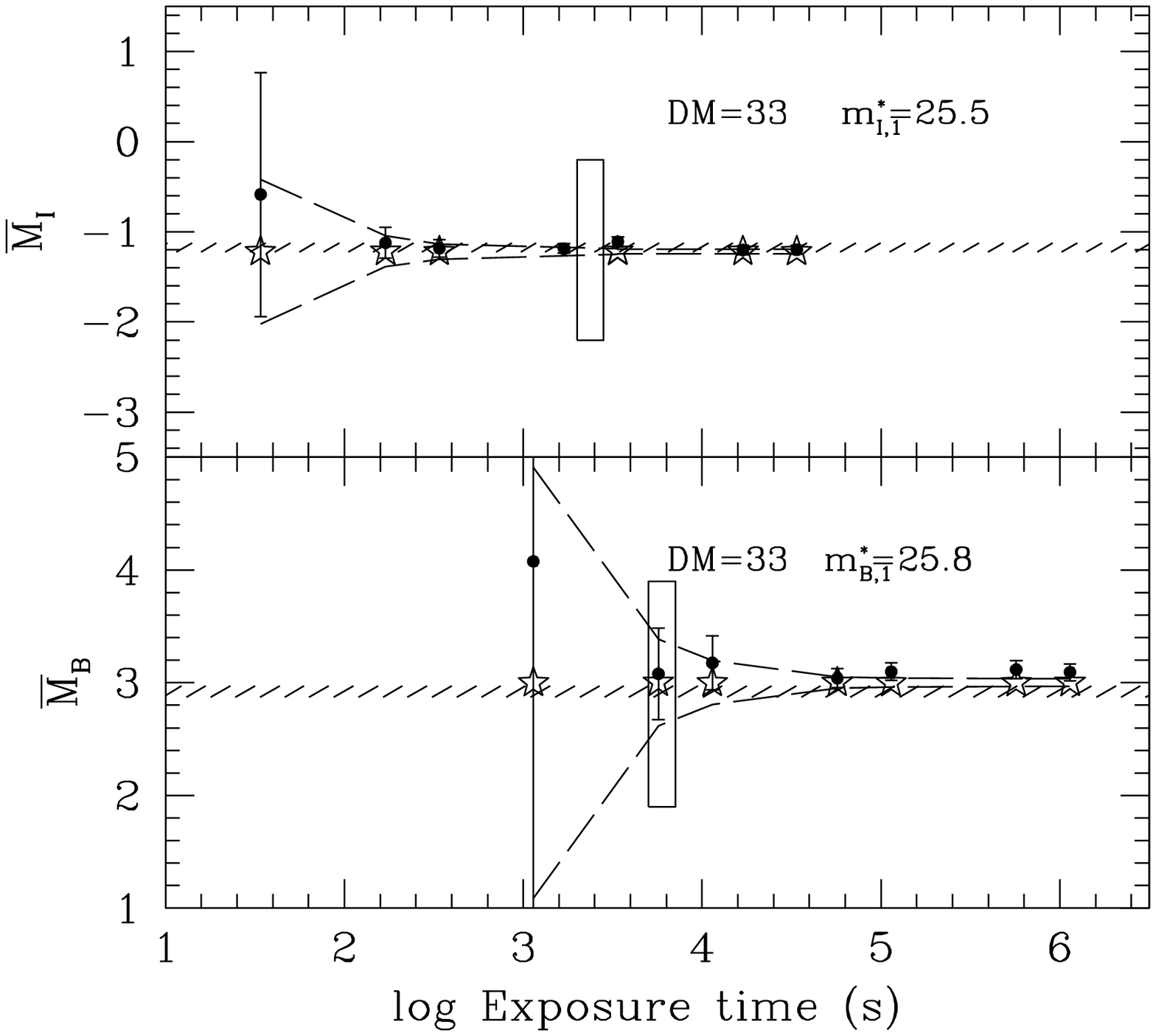}
\figcaption{{\it Left panel } - Absolute SBF magnitudes measured for
the galaxy simulated at distance modulus $\mu_0$=32 against the exposure time adopted for
the simulation. Open stars mark the SBF measured from the frame
without external sources, i.e. the $P_r$ term is zero ($1-\sigma$ area
is also shown with long-dashed lines). Full dots mark the SBF measured
from the final image: it can be recognized that adding external
sources causes a higher dispersion of data and higher
uncertainties at lower exposure times. The shaded horizontal area refers the input SBF
signal. Finally, the boxes locate the positions of our observational
data, for those galaxies at $\mu_0 \sim$32. {\it Right panel} - As left panel
but for $\mu_0$=33.
\label{test2}}
\end{figure}

To check the consistency of the measured SBF signal with the input one
as a function of the S/N ratio of the image, we have performed two
different tests adopting two distance moduli for the galaxy
($\mu_{0}$=32, 33), changing the exposure time in a wide range for
each distance modulus. To properly take into account the role of the
$P_r$ correction, we have assumed the luminosity functions of globular
clusters and background galaxies according to the average properties
of our ACS images. The results of this study are reported in Figure
\ref{test2}.  In the figure we show the input SBF magnitudes (shaded
areas; models are from the Teramo SPoT group\footnote{The Teramo
``Stellar POpulations Tools'' (SPoT) models, from \citet{raimondo05},
are available at the URL: www.oa-teramo.inaf.it/SPoT}), the SBF
measured in the simulated frame before adding external sources (empty
stars), and the SBF magnitudes measured after including also external
sources (filled circles). From these panels can be recognized that the
measured SBF agree with the input signal within uncertainty, but the
uncertainty strongly increases at low exposure times, i.e. low S/N,
since the residual sources $P_r$ and the white noise $P_1$ components
dominate the total fluctuation amplitude.  The properties of the
observational data used for this work are also shown in the Figure
\ref{test2} with boxes located in the regions corresponding to the
proper exposure times (only objects at the correct distance have been
considered for each panel).  In conclusion, these simulations show
that our SBF measurements procedure works well on CCD images having
{\it average} properties similar to our sample of ACS images. However,
it must be pointed out that each single galaxy of the present sample
has different characteristics (amount of external sources, effective
area for the SBF analysis, etc.), which can affect the SBF estimation
in each specific case.

As a further check of the measurements quality, we have obtained SBF
measurements versus the exposure time for real data, by degrading the
original high S/N images to simulate shorter exposure time and smaller
SBF S/N ratio. For this test we have used NGC\,5557 data, splitting
the images of both bands available in three different exposure times:
total exposure time (S/N$\lsim 6$), 2/3 (S/N$\lsim 4$), and 1/2
(S/N$\lsim 3$) of the total exposure time. As a result we find that
the I-band SBF measurements are left practically unchanged, as there
is less than 0.1 mag difference between the two extreme exposure
times. On the contrary, the B-band image with the lowest exposure time
(S/N$< 3$) has too bright SBF, as $\bar{m}_B \sim 34.6 \pm 0.3$
compared to the original $\bar{m}_B \sim 35.2 \pm 0.3$.  Both B- and
I-band SBF amplitudes measured from the 2/3 exposure time frames agree
within uncertainty with the original measurement.

In conclusion, we adopt the value S/N$\sim$3 as limit of separation
between unreliable and reliable SBF measurements for F435W ACS images.
With this choice, we are lead to consider as reliable (within the
quoted uncertainties) the B-band SBF data for NGC\,4696, NGC\,5322 and
NGC\,5557, while NGC\,3258 and NGC\,3268 both lie on the limit of
reliability and NGC\,1407 is below. In the next sections, the B-band
SBF measurement for the last three galaxies will be quoted but not
considered in the analysis and in the discussion.

\section{Discussion}

The common use of SBF magnitudes as a distance indicator relies on the
tight dependence of fluctuations amplitudes on the properties of
stellar populations in galaxies. However, SBF brightness dependence on
stellar population depends on the wavelength interval
considered. Theoretical studies have shown that SBF measured in
filters like B are not good distance indicators because one parameter
(i.e. integrated color) is not sufficient to describe the stellar
population of galaxies, but they are appropriate for stellar
populations analysis. As discussed by several authors
\citep[e.g.,][]{worthey93b,sodemann96,cantiello03,raimondo05}, B-band
SBF magnitudes can be strongly affected by the light coming from hot
luminous stars (extreme HB, Post--AGB, young MS stars, etc.).

The limited number of observational data in the B-band up to now
hampered the ability to check the validity of model SBF
predictions. Furthermore, the available data mainly refer to
galaxies at distances lower than 5 Mpc: M\,32 \citep{sodemann96},
and NGC\,5128 \citep{shopbell93}. These two galaxies will be added
to our sample of objects in the following section. The $\bar{m}_I$
value for NGC\,5128 comes from the \citet{tonry01} database. All
the data adopted for M\,32, and NGC\,5128 are reported in Table
\ref{tab_bi}. The $(B-I)_0$ color for these galaxies is obtained
using the C05 color transformations, upgraded with the new
\citet[][R05]{raimondo05} models.

The measurements presented in this paper increase the sample of B-band
SBF data, and they are the first for a sample of distant giant
elliptical galaxies. Within the limits of the low S/N of the B-band
SBF data, in the following sections we compare B- and I-band SBF
magnitudes and colors with models predictions, in order to point out
the capabilities of blue band SBF as an inquiry tool for stellar
populations, especially in view of applications ($\bar{B}$ radial
gradients, $\bar{B}$-near--IR color data, etc.)  based on future
high-S/N imaging data.

\subsection{Observational properties and comparison with standard SSP models}

In this section we discuss some characteristics of the SBF data for
our sample of objects in the light of most recent and detailed SBF
model predictions.  It is important to emphasize that the forthcoming
comparison is only a ``first approximation'' approach to the general
problem of inferring the physical properties of the stellar
populations generating the SBF signal.

As a first step, we will make use of upgraded and reliable R05
models which -- within the same consistent
theoretical framework -- have been proved to reproduce in detail the
color-magnitude diagrams, the integrated magnitudes and colors and
SBF of well studied systems (e.g. galactic globulars and MC star
clusters) and elliptical galaxies. Even so, one should keep in mind
that SBF predictions include the uncertainties and assumption (IMF,
color transformation, evolutionary tracks, etc.) that typically affect
the theoretical stellar population synthesis models. Thus, taking into
account the observational uncertainties and the small number of
galaxies observed, the present discussion should be considered as
an exploration of the capabilities of SBF method in the B-band more
than a detailed comparison of models with data.

\begin{figure}
\begin{center}
\epsscale{1.}
\plotone{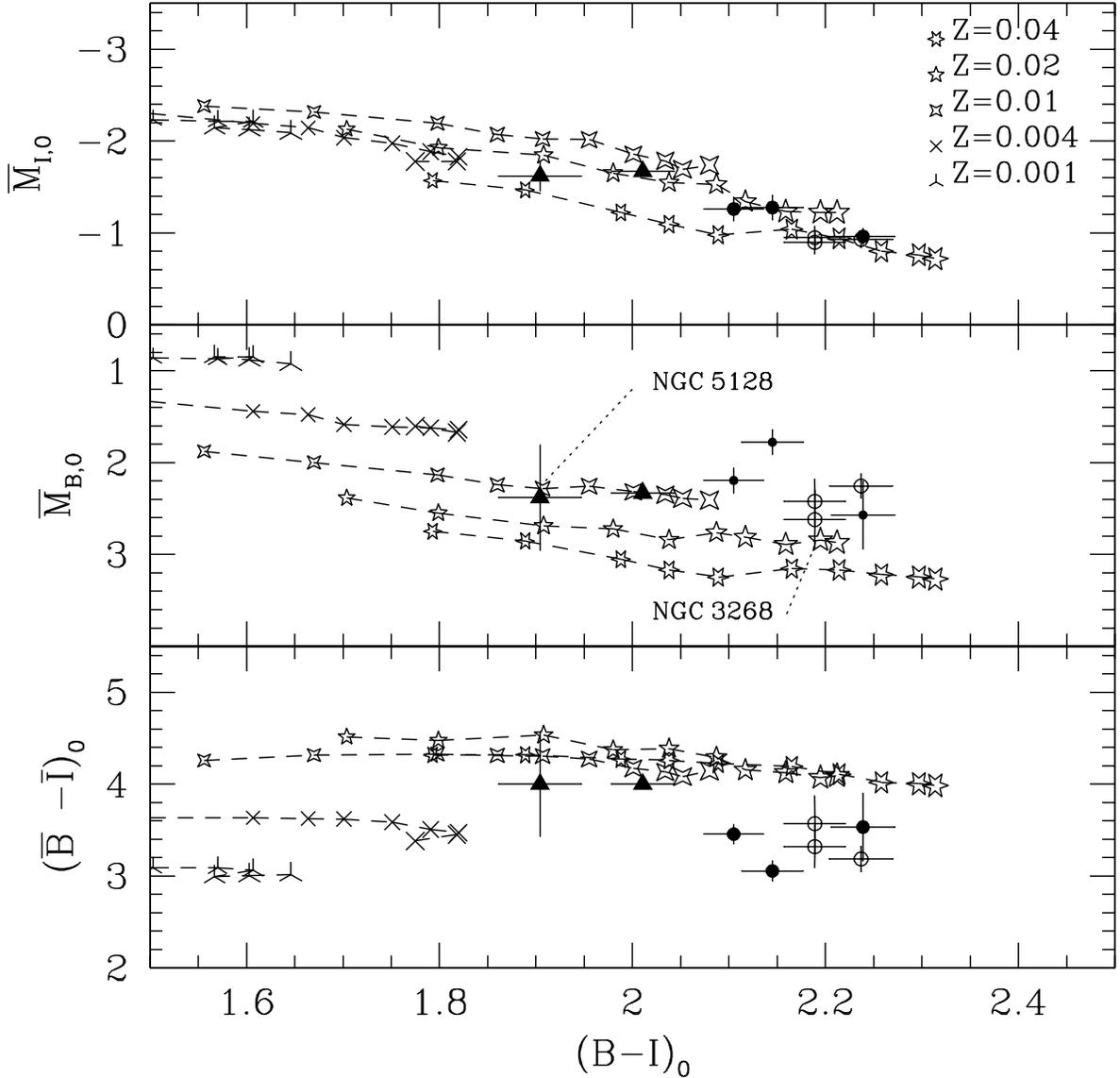}
\figcaption{Observational data compared with models for various
chemical compositions (upper right labels), with ages 1.5, 2, 3, 4, 5,
7, 9, 11, 13, and 14 Gyr (symbols with increasing size mark models of older
age). Models are from SPoT web site (R05). Full (empty) circles mark
the location of our ACS B-band reliable (unreliable) SBF measurements.
Filled triangles show the location of M\,32 and NGC\,5128.
M\,32 and NGC\,5128 $(B-I)_0$ error bars are evaluated taking into account
the uncertainties of the $(B-I)_0$-to-$(V-I)_0$ color transformation.
\label{spot}}
\end{center}
\end{figure}

In Figure \ref{spot} we compare SBF and color data with the recent R05
SSP models. The upper two panels in Figure \ref{spot} show the
$\bar{M}_I$ and $\bar{M}_B$ versus the galaxy integrated color.
Absolute SBF magnitudes are derived using the distance moduli reported
in Table \ref{tab_bi}. SBF color data and models are shown in the
lower panel. In this Figure we plot with different symbols the
reliable data (full circles), and the B-band unreliable data (empty
circles).  Also the data of M\,32 and NGC\,5128 are reported (filled
triangles).  For completeness and, additionally, to emphasize possible
inhomogeneities emerging between the different bands, in the following
paragraphs we will analyze the galaxy properties emerging from each
one of the panels in Figure \ref{spot}.

\begin{itemize}

\item {$\bar{M}_I$ versus $(B-I)_0$ } -- I-band comparison of the
models to the data shows excellent agreement of R05 SBF models
with SBF measurements. This result substantially support the
finding presented in C05, even though here we are discussing one
single averaged SBF measurement, instead of SBF gradients.
Specifically, an old t$\gsim$10 Gyr, metal rich Z$\gsim$0.02
stellar population dominates the light emitted by these galaxies.

For the case of M\,32, SBF data are well reproduced by R05 SSP
models with an age $\sim$5 Gyr and $Z \lsim$ 0.02. This result is
very similar to what found by \citet{cantiello03}, and by other
authors \citep[e.g.,][]{trager00b} from line strength analysis.

NGC\,5128 (Centaurus A) data are consistent with SSP models having
$t\gsim 3$ Gyr, and $Z \lsim$ 0.02. The average metallicity inferred
from resolved halo stars is $Z \gsim 0.005$, with a large spread; two
age peaks are recognized, one at 2 Gyr and one at older ages
\citep{marleau00}.  Moreover, the complexity of the NGC\,5128 stellar
system increases as one approaches the regions of the dark absorption
lanes in the galaxy \citep{rejkuba01}. Note that the resolved stars
data refer to regions not overlapping with the SBF data measured
closer to the center of the galaxy. Broad band colors in regions
overlapping ours have been derived by several authors
\citep{vandenbergh76,dufour79}. These authors settled the twofold
character of the NGC\,5128 stellar populations, with a main body
component which is old but bluer than usual in elliptical galaxies,
consistent with a 7-9 Gyr, Z$\sim 0.01$ stellar system, and a disk of
young, metal rich stars. In conclusion, taking into account the
complexity of the stellar system of NGC\,5128, and the integrated
nature of the SBF signal data, we consider as satisfactory the
agreement between the properties of stellar systems inferred from SBF
versus color data/models comparisons and properties taken from
literature.

\item{$\bar{M}_B$ versus $(B-I)_0$ } -- Inspecting the middle panel of
Figure \ref{spot} we find that the B-band SBF magnitudes are brighter
than the model predictions, in contrast to the good agreement obtained
with the I-band. The only exception being NGC\,5128 (and NGC\,3268
which, however, we consider at the limit of reliability) due to the
big error bars.  All the galaxies at $(B-I)_0 \gsim 2.1$ mag appear to
have substantially brighter SBF magnitudes respect to models
predictions.

Also in the case of M\,32 the age and chemical composition derived
from this panel do not agree with properties inferred from I-band SBF
or literature data.  Again, the SBF measured is brighter respect to models
expectations for a t$\sim$5 Gyr Z$\sim$0.02 stellar system.

\item{$(\bar{B}-\bar{I})$ versus $(B-I)_0$} -- The additional distance
modulus uncertainty present in the top two panels of Figure
\ref{spot}, is removed in the bottom SBF-color versus color panel.
However, as shown in this panel, no substantial improvement occurs by
using the distance-free SBF-color versus integrated color with respect
to the disagreement presented before.

\end{itemize}

\begin{figure}
\begin{center}
\epsscale{1.2}
\plottwo{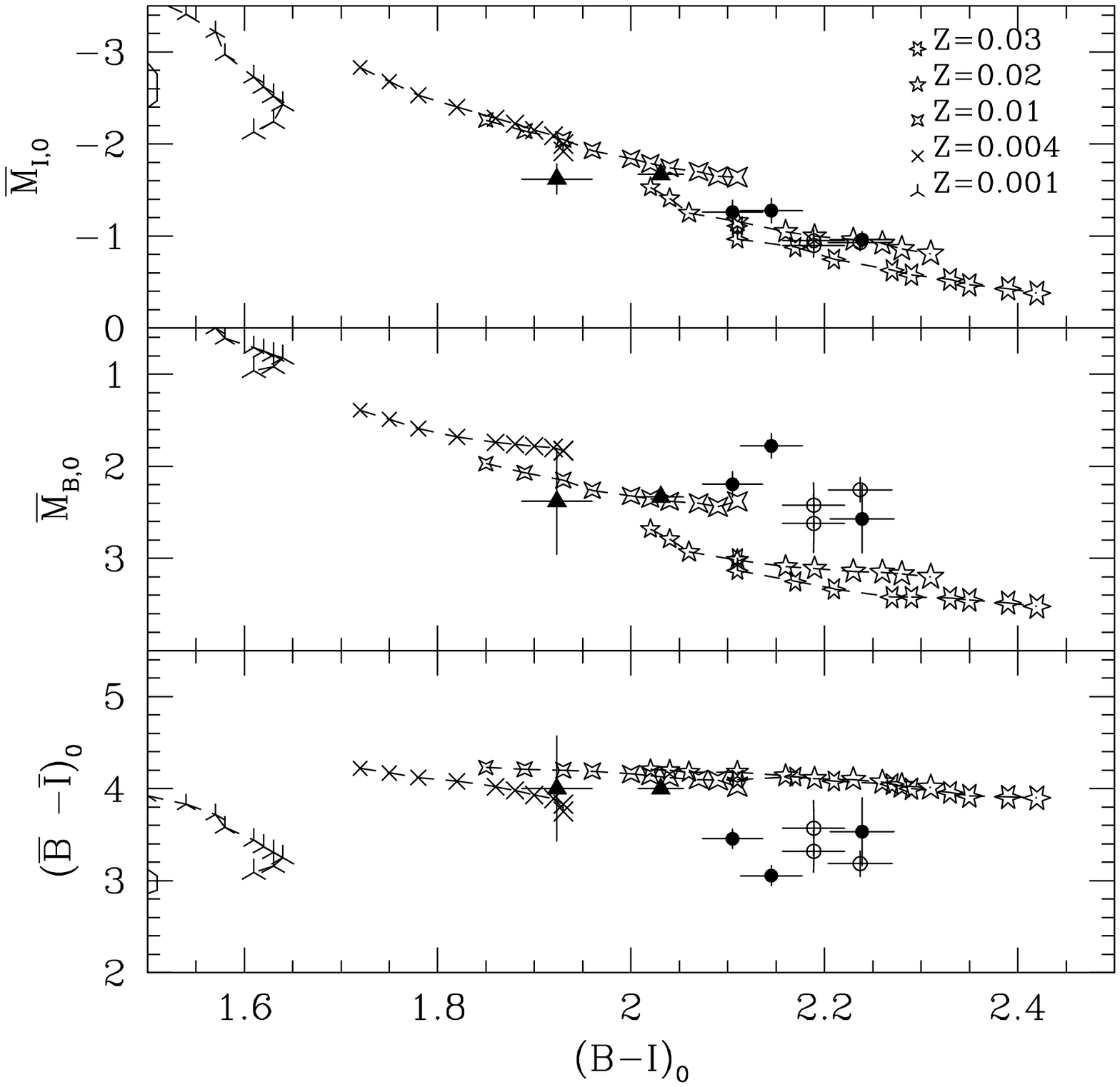}{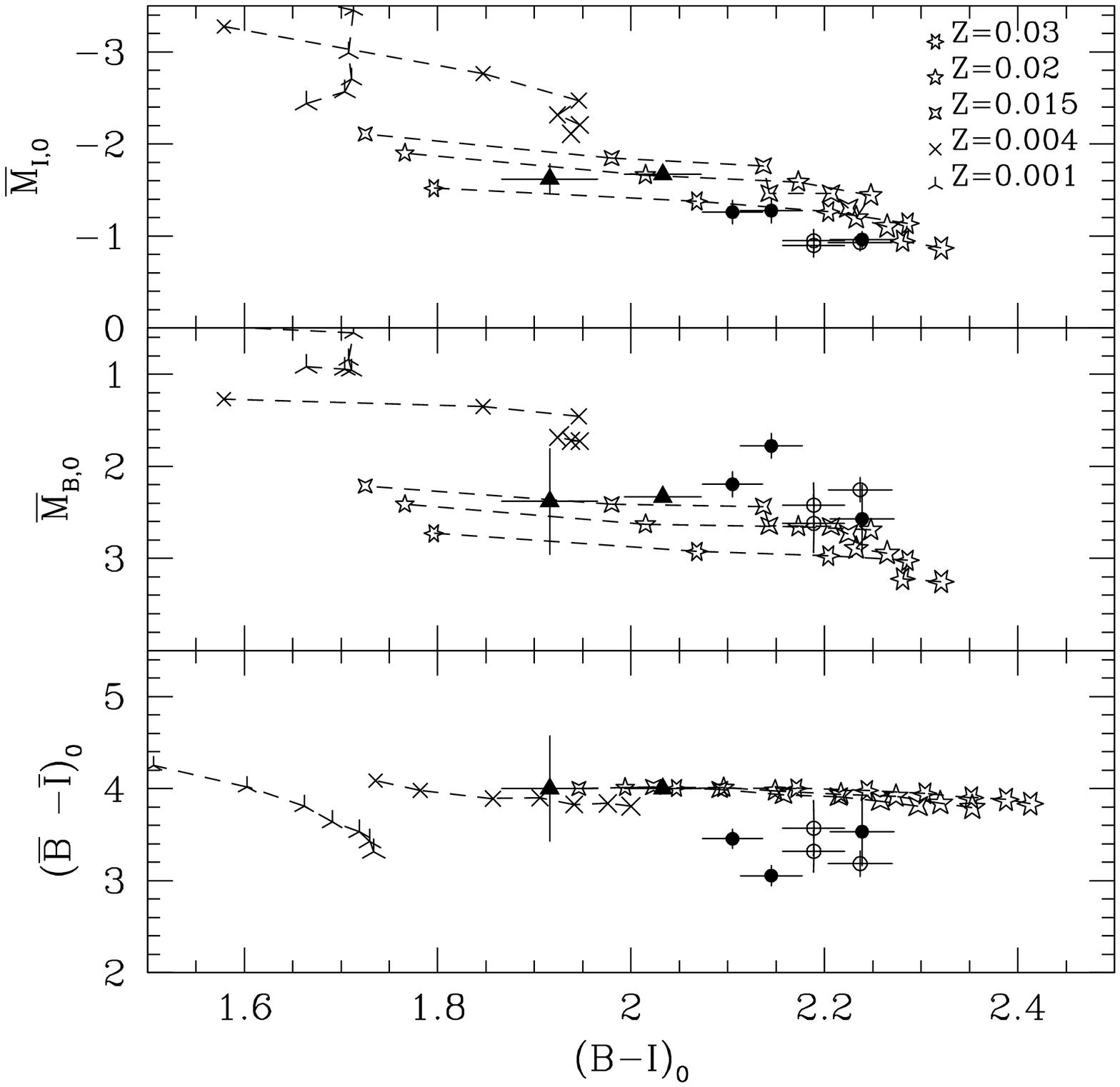}
\figcaption{Same as Fig. \ref{spot} but for \citet[][upper panel]{bva01}, and
\citet[][lower panel]{marin06} models. The age of upper panel models ranges from
4 to 18 Gyr, with an increasing step of $\sim12$\%.
The ages for lower panel models are 3, 5, 7, 9, 11 and 13 Gyr \citep[stellar tracks from][]{bertelli94}.
Chemical compositions symbols are labeled upper right in the panels for both set of models.
\label{others}}
\end{center}
\end{figure}

To analyze the possible origin of the mismatch when the B-band data
are considered, we have compared our SBF and color data with other
sets of stellar population models \citep[][Figure \ref{others} shows
the latter two set of models]{worthey94,bva01,marin06}. However,
adopting different SBF models does not solve the mismatch in the
B-band.  This is quite interesting because it can be interpreted
(within the quoted small statistics and wide observational error
bars) as an indication of a missing contributor to the SBF signal in
the canonical B-band SSP models.
A further possibility is the eventuality of a systematic bias of the
B-band measurements which may explain the disagreement without
requiring implication of the SSP models. On the basis of extended
experiments with simulations we are leaded to discard the presence
of systematic offsets in the B-band data. However, this hypothesis
could not be ruled out due to the small statistics of our sample.

\subsection{Comparison with non-standard SSP models}

A possible alternative to understand such B-band data/models
disagreement is to consider non standard stellar population
models. As examples of non standard models, we take into account
two different cases: $(i)$  SSP models with a non standard ratio
of Post--AGB to HB number of stars; $(ii)$ SBF models of Composite
Stellar Populations (CSP).

$(i)$ Following the discussion concerning the contribution to SSP of
hot stars experiencing bright and fast evolutionary phases presented
 by \citet{brocato90,brocato00}, and taking into account the
\citet{worthey93b} comments on the effects of these stars on SBF
magnitudes, in \citet{cantiello03} we presented a detailed analysis
of the SBF magnitudes against the number of hot stars in the late
evolutionary phases (Post-AGB). In that work it is shown that SSP
models with an increased ratio of the number of Post-AGB stars with
respect to the number of HB stars ($N_{Post-AGB}/N_{HB}$), have
brighter $\bar{M}_B$ amplitudes, while the $\bar{M}_I$ and colors
are left practically unchanged.

\begin{figure}
\begin{center}
\epsscale{1.}
\plotone{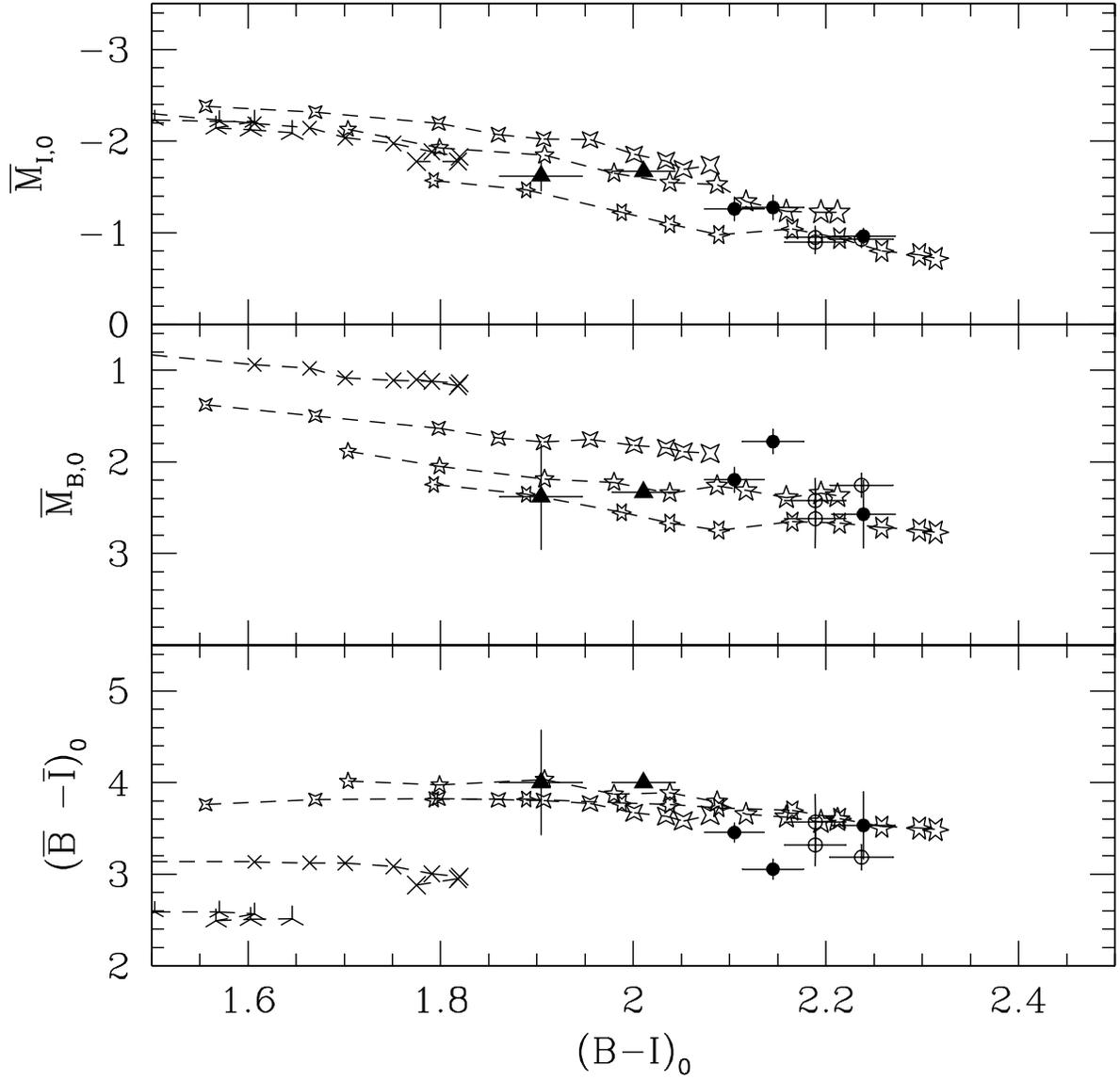}
\figcaption{Same as Fig. \ref{spot} but models with an increased
$N_{Post-AGB}/N_{HB}$ ratio (see text) are plotted.
\label{hot}}
\end{center}
\end{figure}

In Figure \ref{hot} we show the comparison of the models to the data
adopting SSP models with a $N_{Post-AGB}/N_{HB}$ ratio doubled with
respect to standard models. As can be recognized from the Figure, all
the stellar populations properties inferred from the B-band models
agree, within the observational error bars, with the same I-band
models described above, except for the NGC\,5557 data.  This means
that SSP models which include a component of hot, bright stars
undetectable in the SBF I-band models, but with a non negligible
effect on B-band SBF measurements, is able to reconcile the mismatch
we find between standard SSP models and B-band measurements.

$(ii)$ B-band SBF amplitudes may reveal the presence of populations of
hot stars (extreme HB stars, post-AGB stars, young populations,
metal-poor stars, etc.), even when they only represent a small
fraction of the overall stellar population.  B-band SBFs are therefore
a valuable tool to investigate Composite Stellar Population. As shown
by C05, in some cases the combination of a dominant stellar population
with secondary hot components of different chemical compositions
and/or ages, eventually allows to better explain some observational
properties of the galaxy.  For this comparison we adopt the
\citet[][BVA01 hereafter]{bva01} CSP models, thus our conclusions must
be considered as valid only within the limits of such CSP scenario.
The BVA01 CSP models are obtained combining homogeneous SSP models in
such a way as to mimic, at least approximately, the evolution of an
elliptical galaxy. We refer the reader to the BVA01 paper for a
detailed description of their composite models. Briefly: SSP models
are grouped into three bins according to metallicity: metal-poor
($0.0004\leq Z \leq 0.001$, m.p. hereafter), intermediate ($0.004\leq
Z \leq 0.008$, int.), and metal-rich ($0.02\leq Z \leq 0.03$,
m.r.). Then one SSP model is randomly chosen from each metallicity
bin, with some age restrictions depending on the bin.  The three
components are then combined according to random weighting factors
[$f_{m.p.}$, $f_{int.}$, $f_{m.r.}$], and fluctuation amplitudes
calculated using a generalization of the \citeauthor{ts88} formula.

\begin{figure}
\begin{center}
\epsscale{1.}
\plotone{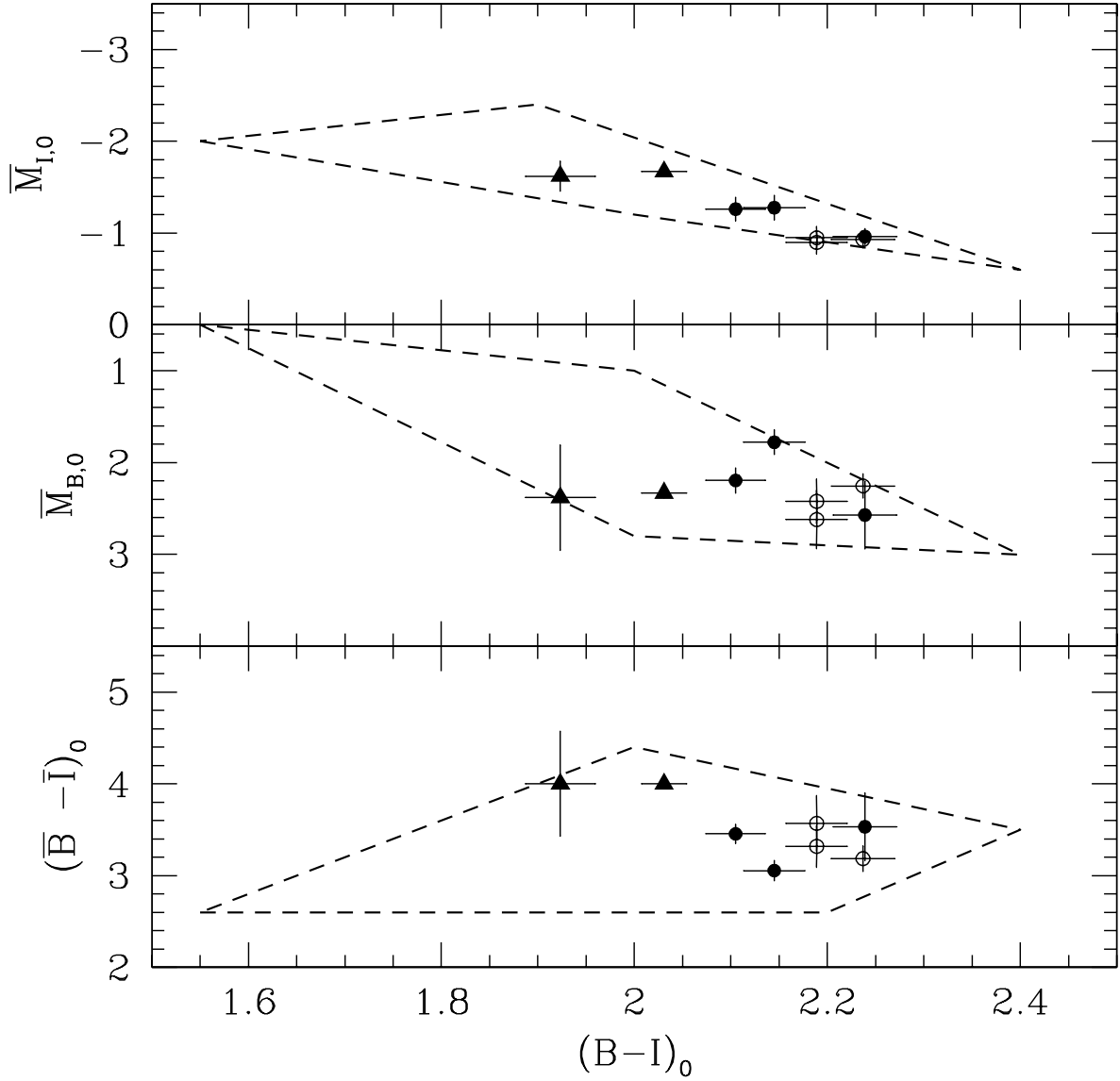}
\figcaption{SBF data are compared with CSP models from BVA01.
Symbols for observational data are as in Figure \ref{spot}.
For sake of clearness, only the edges of the CSP locations are shown.
\label{compos}}
\end{center}
\end{figure}

In Figure \ref{compos} the comparison of these models with
observational data is presented. In the Figure we mark the edges of
the area covered by models. As shown in the panels of the Figure, the
presence of CSP eliminates the problems existing with previous
standard SSP models, as all data lie within the area of the models.
By inspecting the average properties of CSP models overlapping with
observational data, one concludes that:

\begin{itemize}

\item the galaxies at $(\bar{B}-\bar{I})_0\sim 3.4$ mag and
  $(B-I)_0\sim 2.2$ mag are strongly dominated by an old (t$\gsim$14
  Gyr) metal rich stellar system, with a possible minor contribution
  due to an intermediate metallicity stellar component ($f_{int.}\lsim
  20$\%), and a negligible amount of a metal-poor component
  ($f_{m.p.}\lsim 5$\%)

\item M\,32 data agree with CSP models composed by a comparable
fraction of a metal-rich stellar system of 5-7 Gyr, and an
intermediate metallicity population with t$\sim$ 11 Gyr.

\item NGC\,5128 is the only case where the dominant component is not
the metal-rich one. In this case a t$\sim$9 Gyr, intermediate
metallicity stellar system appears to be the dominant stellar
component ($f_{int.}\sim$60\%), with a secondary t$\sim$7 Gyr
metal-rich one.

\item for NGC\,5322 observational data overlap with models having
$\sim$70\% of light coming from a metal-rich t$\sim$ 13 Gyr component,
and a substantial fraction of light coming from an intermediate
metallicity ($f_{int.}\sim$20\%), and a metal-poor ($f_{m.p.}\sim$
10\%) component.  A similar result is found for NGC\,5557, with the
difference that the metallicity of the most metal poor component in
this case is Z$\sim$0.0004.

\end{itemize}

In addition to the two previous non-standard SSP models, we have also
considered the case SSP obtained using $\alpha$-enhanced stellar
tracks, based on the recent models by \citet{lee06}. The use of
$\alpha$-enhanced SBF/color models lead to estimate chemical
compositions which are on average a factor 0.3 dex higher respect to
our previous conclusions, with substantially unchanged ages, but the
mismatch aforementioned is not reconciled. Thus we can exclude the
$\alpha$-enhancement as the driving source of the B-band mismatch, at
least within the limit of the scenario presented by \citet{lee06}.

Although the above conclusions must not be over interpreted and
considered as valid only within the limits of the specific
non-standard models taken into account and the low S/N of the B-band
SBF data, these results possibly point out that ($i$) I-band SBF
magnitudes are confirmed as a powerful distance indicator, as they
exhibit a small dependence on the detailed properties of the stellar
population in the galaxy; ($ii$) B-band SBF amplitudes are sensibly
affected by also a small component of a hot stellar system (extreme
HB stars, Post-AGB stars, young populations, metal-poor components,
etc.) and represent a valuable tool to study the presence of these
stars and/or stellar populations in unresolved stellar systems.

\section{Conclusions}

In this paper we have carried out the first extensive study on
B-band SBF measurements for distant galaxies, based on
measurements obtained from ACS imaging data for a sample of six
elliptical galaxies. In a previous paper (C05) we succeeded in
detecting I-band SBF radial gradients for the same sample of
objects. The quality of B-band images did not allow us to obtain
SBF measurements in different galaxy regions. Moreover, after a
few image tests, we have decided that only three of these B-band
SBF measurements could be considered as reliable.

We have added to our sample the only two other galaxies with I-
and B-band SBF measurements available from literature: M\,32 and
NGC\,5128. The analysis of the observational properties of this
sample of galaxies shows that using the standard SSP models from
the SPoT group, and various models from other authors, the stellar
population properties derived from I-band SBF versus $(B-I)_0$
color data/models comparison agree with the stellar population
properties known in literature derived from other indicators.
Briefly: the most massive objects appear to be dominated by an old
(t$\gsim$ 10 Gyr) metal-rich Z$\gsim$0.02 stellar component; M\,32
light is dominated by a t$\sim$5 Gyr Z$\sim$0.02 stellar system,
while NGC\,5128 light seems to be dominated by a t$\sim$ 3 Gyr,
Z$\gsim$ 0.01 population.

In spite of this, we find that the B-band comparison of the models
to the data is not satisfactory, especially for the red [$(B-I)_0
\gsim 2.1$] galaxies in our sample. In order to try to solve such
disagreement we have presented a comparison of data with non
standard stellar population models.  In particular we have used
prescriptions based on two different approaches. ($i$) Models with
an enhanced number of hot (Post-AGB) stars: these models appear to
resolve almost completely the disagreement present with the standard
SSP models, within the limit of the observational uncertainty. The
only galaxy whose properties do not seem to be well interpreted in
this scenario is NGC\,5557. ($ii$) Composite Stellar Population
models: within this scenario the SBF measurements are all included
in the region of the SBF-color diagrams covered by the CSP models.
In particular, these models show that the light of the galaxies in
our sample at $(B-I)_0 \gsim 2.1$ mag can be interpreted as
dominated by an old, t$\gsim$10 Gyr, Z$\gsim$0.02 stellar component,
in some cases accompanied by a non negligible amount of lower
metallicity stellar components. As an example, NGC\,5557 data are
nicely reproduced by models including a small fraction ($\sim$10\%
of the total light) of old Z$\sim$0.0004 stars.

In conclusion, our results confirm the theoretical expectations that
$\bar{B}$ data are not well suited for distance study and, together
with SBF colors like the $\bar{B}-\bar{I}$, they represent a valuable
tool to investigate the properties of unresolved stellar systems, with
particular regard to the hot stellar component.  As a concluding
remark, we emphasize that the use of B-band SBF, SBF-color, and SBF
radial gradients data to study the properties of the stellar systems
of external galaxies, rely on the availability of high quality
observational data. The data used for this work are generally
characterized by low S/N ratios, based an the simulations and results
presented in this paper for B-band SBF, we suggest that higher S/N
data (S/N$\gsim$10) is required to carry out a detailed study of
stellar population properties based on this technique.

\acknowledgements
Financial support for this work was provided by COFIN 2004,
under the scientific project ``Stellar Evolution'' (P.I.: Massimo
Capaccioli).

\appendix
\section{Simulating ``realistic'' CCD images of elliptical galaxies}

In this appendix we outline the major steps of the procedure
to simulate CCD images of elliptical galaxies adopted in section 3.
The procedure has been properly developed to include the SBF signal
in the galaxy model.

One of the basic inputs of these simulations is the R05 SSP
models, which fix the chemical content and the age of the SSP for
each simulation.  One of the capabilities of the SPoT stellar
synthesis code is to generate $N_{sim}$ independent simulations of
SSP with fixed age and metallicity.  Once the SSP total mass is
fixed, the SPoT code is able to randomly populate the IMF of the
stellar system. The code is implemented to reliably simulate stars
on all post-MS phases: RGB, HB, AGB, etc.; see
\citet{brocato99,brocato00} and \citet{raimondo05} for the
details. At fixed age, metallicity, and IMF shape, the simulations
obtained have, on average, similar properties (according to the
average mass of the population), but each single simulation
differs from the others for statistical reasons.

We use such capability $(1)$ to simulate a ``realistic'' galaxy,
which also includes the SBF signal, whose amplitude is fixed by
the input properties of the SSP. Then, $(2)$ we measure
fluctuations amplitudes of the simulated galaxy applying the
procedure described in section 2.2.3. Finally, $(3)$ the SBF
measured is compared with the input one to verify if and how, in
the case of low S/N images, the measurements are affected.

First of all, we simulate a galaxy with a \citet{devaucouleurs48}
$r^{1/4}$ luminosity profile, although the procedure accepts also
a generic \citet{sersic68} $r^{1/n}$ profile. Starting from the
smooth {\it analytic profile} whose SBF is zero, we built a {\it
realistic profile} to simulate the poissonian fluctuation due the
star counts which, as mentioned before, constitutes the physical
basis of the SBF signal. To simulate the fluctuations, at each
fixed radius $r_*$, we substitute the well defined analytic
surface brightness $\mu(r_*)_{an}$ in all $N_{pix}$ pixels
corresponding to this radius, with the surface brightness of a
simulated SSP belonging to a sample of $N_{pix}$ SSP simulations
having $\langle \mu(r_*)_{SSP} \rangle = \mu(r_*)_{an}$, in this
way the brightness profile of the galaxy is preserved and the
poissonian fluctuation due to star count fluctuation is included.

This procedure (which is repeated at all radii) requires a large
number of SSP models, according to the number or radii one wants
to simulate. We note once more that, since the SSP is chosen by
the user, the galaxy SBF amplitude is an input parameter; for
example by using the R05 models for an SSP having Z=0.02 and
$t=14~Gyr$, the input SBF is $\bar{M}_B\sim 2.9$ in B-band, or
$\bar{M}_I\sim -1.2$ in case of I-band simulations.

\begin{figure}
\epsscale{1.}
\plottwo{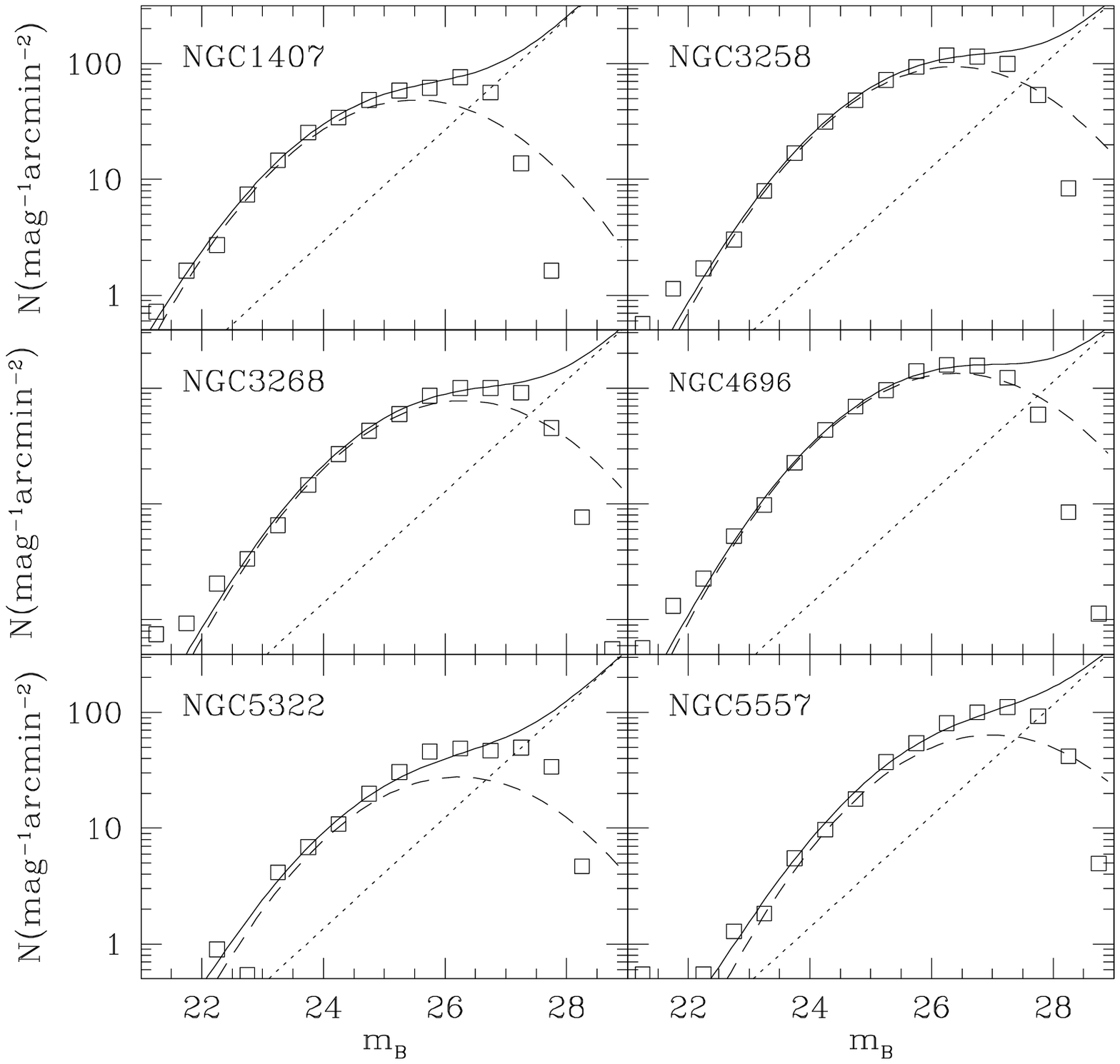}{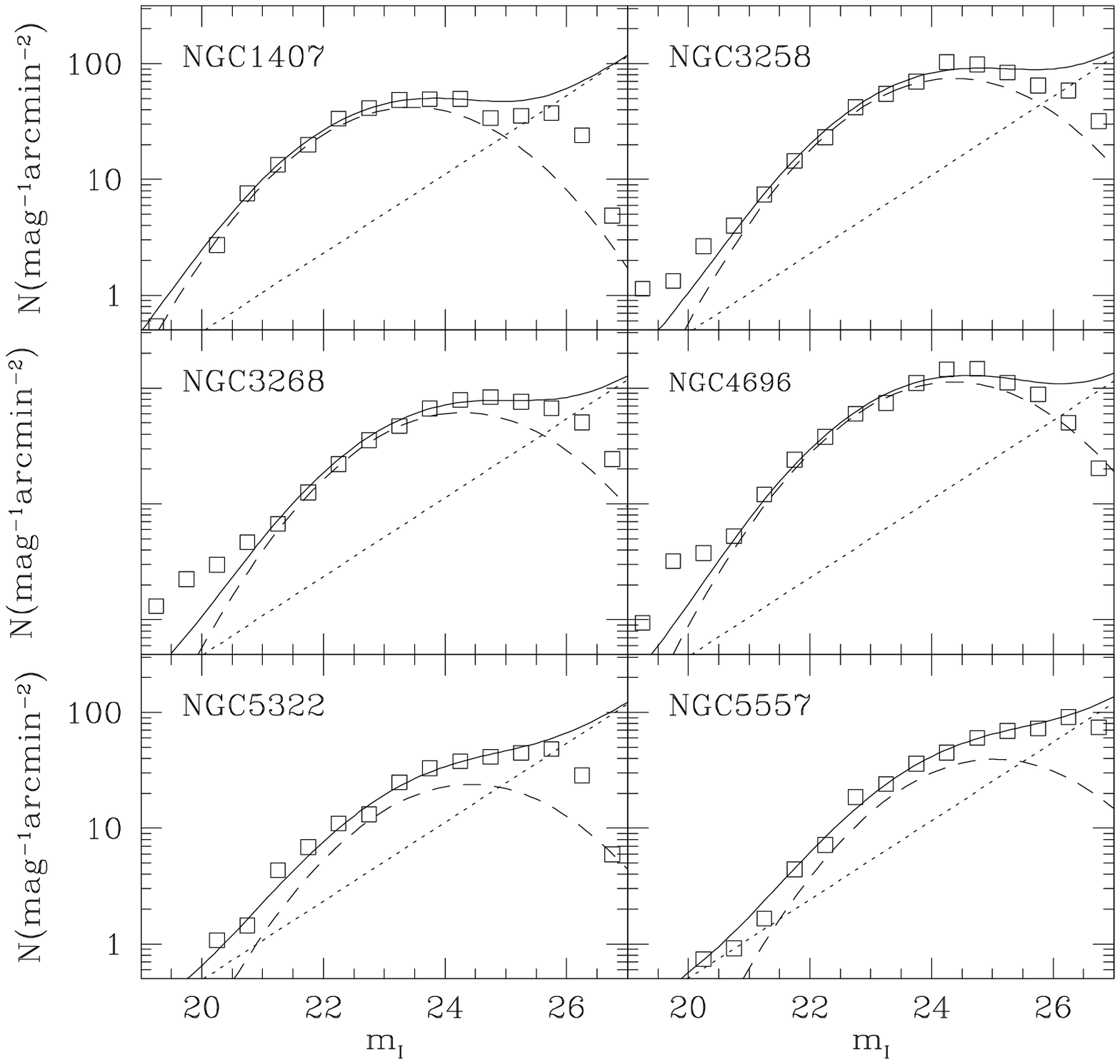}
\figcaption{The B-band (left panels) and I-band (right panels)
total luminosity function of external sources.
The panels show the observed total number density as a function of
the magnitude (open squares), and the model luminosity function
used to estimate the residual contribution to fluctuations
arising from undetected sources (solid line). The two components
of the model number density, globular clusters and background galaxies,
are also shown with dashed and dotted lines, respectively.
\label{lumfunc}}
\end{figure}

At this point, to simulate fully realistic CCD image, we included the
effect of the PSF, external sources (globular clusters and galaxies),
and instrumental noise. To simulate the effect of the PSF on the data,
the galaxy image is convolved with the point spread function profile
before adding the external sources (which are already convolved with
the instrumental PSF). The external sources were added according to a
fixed total luminosity function of globular clusters and
galaxies. This luminosity function, similar to those shown in Figure
\ref{lumfunc}, is of the sum of a power law for galaxies, and a
GCLF. Galaxies are randomly distributed on the galaxy image, Globular
Clusters were distributed using an inverse power law centered on the
galaxy.  A constant sky value is also added to the image.  Finally the
photon and detector noises are added by using the IRAF task {\it
mknoise}, according to the readout-noise and gain properties of ACS.

It must be emphasized that, in addition to the input parameters
already introduced (galaxy profile index, density and spatial
distribution of GCs, density of background galaxies, PSF shape,
age and metallicity of the SSP), there are also some other
user-defined parameters  in the galaxy simulation which are not
discussed here (distance, effective magnitude and effective
radius, exposure time, field of view, zero point magnitude, sky
brightness, etc.).

Once the simulation has been completed, the whole procedure described
in section 2.2.3 (sky evaluation, galaxy modeling, sources detection
and masking, LF fitting, SBF measurement) is run on the simulated CCD
images.  

\begin{figure}
\begin{center}
\epsscale{1.} \plottwo{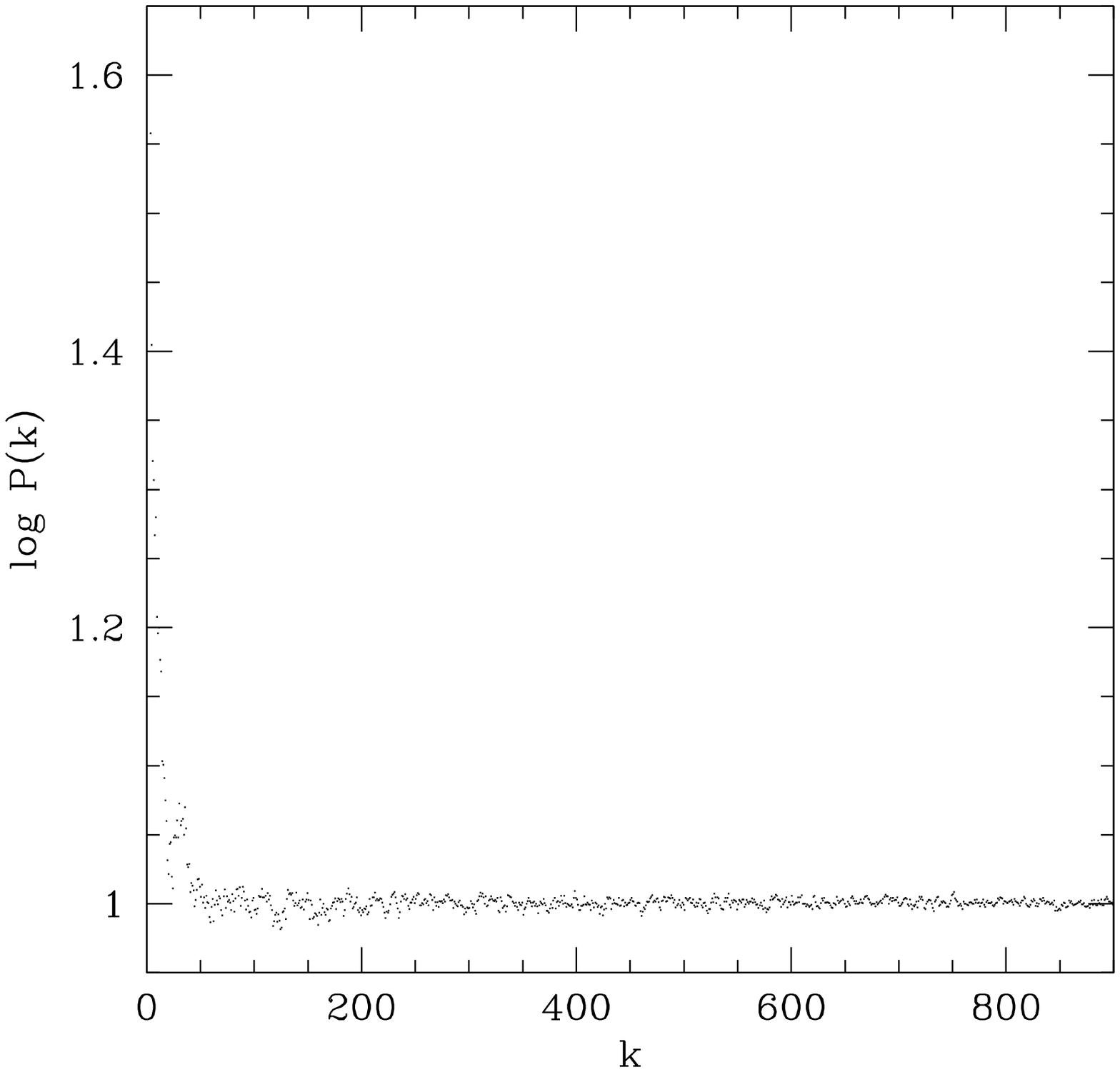}{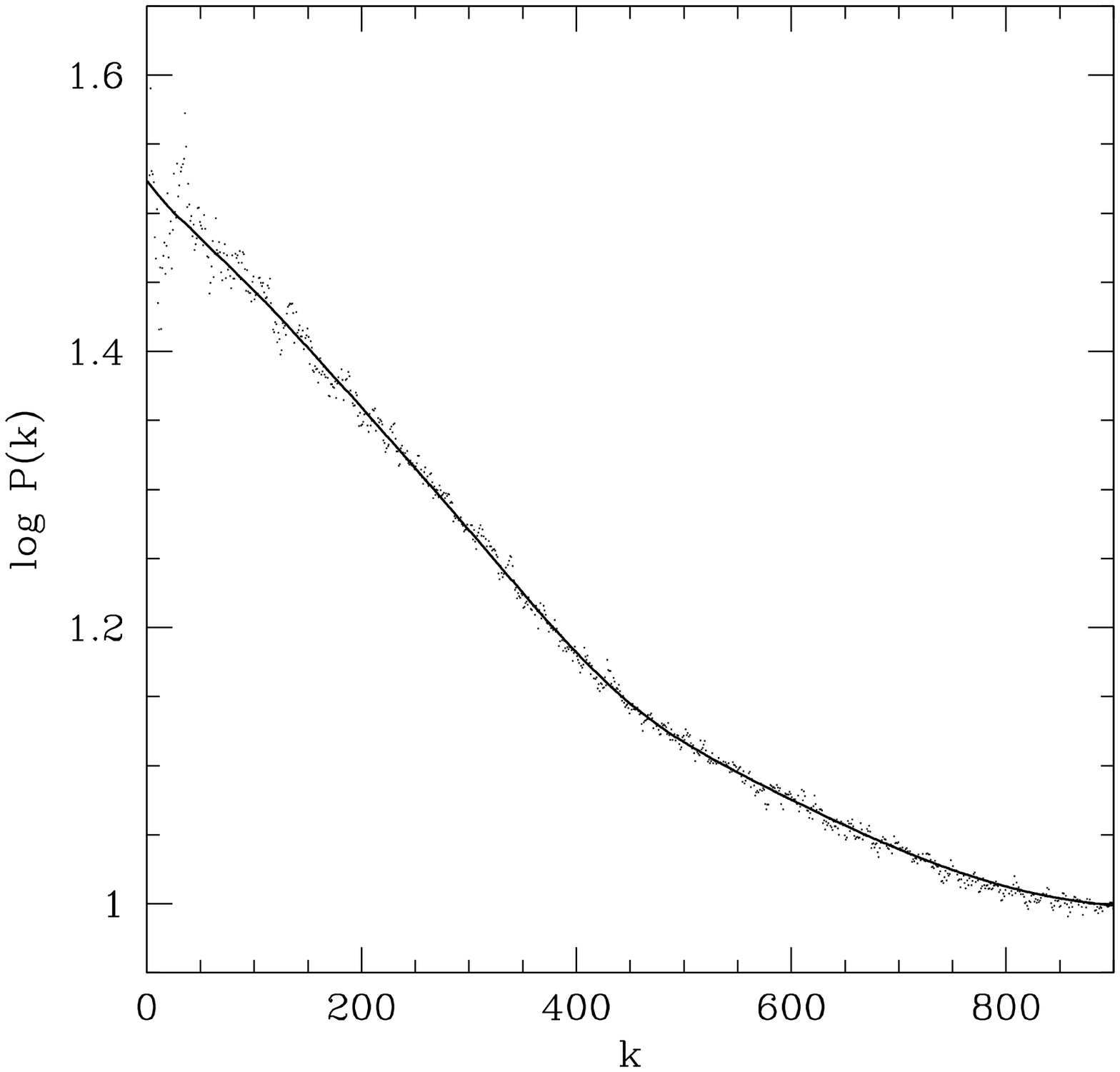} \figcaption{Power spectra
of the residual frame for galaxy simulated without (left panel)
and with (right panel) the SBF pixel-to-pixel variation. The
strong variations visible at $k\leq 50$ are due to the extra
correlation at these scales added by the large scale smoothing
operation. In the right panel, the best fit to the data is shown
as a solid line. \label{test1a}}
\end{center}
\end{figure}

In order to show the effect of introducing the Poissonian variation
between adjacent pixels, in Figure \ref{test1a} we show the power
spectrum of the residual frames for an image simulated with (right
panel), and without (left panel) the SBF signal.  The difference
clearly emerges: the spectrum of the image without SBF is flat (i.e.,
no SBF signal is revealed), while the power spectrum of the image with
SBF has the characteristic PSF shape.

\acknowledgements We thank the anonymous referee for helping us to
improve this paper with constructive criticisms and useful
suggestions.

%%%%%%%%%%%%%%%%%%%%%%%%%%%%%%%%%%%%%%%%%%%%%%%%%%%%%%%
%%%%%%%%%%%BIBLIO %%%%%%%%%%%%%%%%%%%%%%%%%%%%%%%%%%%%%
%%%%%%%%%%%%%%%%%%%%%%%%%%%%%%%%%%%%%%%%%%%%%%%%%%%%%%%

\bibliographystyle{apj}
\bibliography{cantiello}

%%%%%%%%%%%%%%%%%%%%%%%%%%%%%%%%%%%%%%%%%%%%%%%%%%%%%%%
%%%%%%%%%%%TABLES %%%%%%%%%%%%%%%%%%%%%%%%%%%%%%%%%%%%%
%%%%%%%%%%%%%%%%%%%%%%%%%%%%%%%%%%%%%%%%%%%%%%%%%%%%%%%
\clearpage

\begin{deluxetable}{lcccccccc}
\tablewidth{0pt}
\tabletypesize{\scriptsize}
\tablecaption{Observational Data}
\startdata
\hline
\multicolumn{1}{c}{Galaxy} & \multicolumn{1}{c}{R.A.} & \multicolumn{1}{c}{Decl.} &  
\multicolumn{1}{c}{$v_{cmb}$} & \multicolumn{1}{c}{T} & \multicolumn{1}{c}{$A_B$} & 
\multicolumn{1}{c}{F814W} & \multicolumn{1}{c}{F435W}  \\
\multicolumn{1}{c}{} & \multicolumn{1}{c}{} & \multicolumn{1}{c}{} & 
\multicolumn{1}{c}{($km/s$)} & \multicolumn{1}{c}{} & \multicolumn{1}{c}{} & \multicolumn{1}{c}{exp. time($s$)} & 
\multicolumn{1}{c}{exp. time ($s$)} & \multicolumn{1}{c}{} \\
\multicolumn{1}{c}{(1)} & \multicolumn{1}{c}{(2)} & \multicolumn{1}{c}{(3)} & 
\multicolumn{1}{c}{(4)} & \multicolumn{1}{c}{(5)} & \multicolumn{1}{c}{(6)} & \multicolumn{1}{c}{(7)} &
\multicolumn{1}{c}{(8)} \\
\hline
NGC\,1407 &  55.052 & -18.581 & 1627 & -5  & 0.297 & 680  & 1500 \\
NGC\,3258 & 157.226 & -35.606 & 3129 & -5  & 0.363 & 2280 & 5360 \\
NGC\,3268 & 157.503 & -35.325 & 3084 & -5  & 0.444 & 2280 & 5360 \\
NGC\,4696 & 192.208 & -41.311 & 3248 & -4  & 0.489 & 2320 & 5440 \\
NGC\,5322 & 207.315 & 60.191  & 1916 & -5  & 0.061 & 820  & 3390 \\
NGC\,5557 & 214.605 & 36.494  & 3433 & -5  & 0.025 & 2400 & 5260 \\
\enddata
\tablenotetext{Notes}{Col. (1): Galaxy name; Cols. (2) and (3): Right
  ascension and declination from RC3; Col. (4): Recession velocity in
  the CMB reference frame from \citet{tonry01}; Col. (5): Morphological T-type from RC3;
  Col. (6): B-band extinction from \citet{sfd98};
Col. (7) and (8): total exposure time for F814W and F435 images, respectively.}
\label{tab_data}
\end{deluxetable}

\begin{deluxetable}{cccccccccccc}
\tabletypesize{\scriptsize}
\rotate
\tablecaption{Fitting Parameters}
\tablewidth{0pt}
\startdata
\hline
\multicolumn{2}{c}{} & \multicolumn{5}{c}{B-band fitting parameters (ADU/exposure)} & \multicolumn{5}{c}{I-band fitting parameters (ADU/exposure)} \\
\multicolumn{1}{c}{Galaxy} & \multicolumn{1}{c}{r(arcsec)} &\multicolumn{1}{c}{$P_0$} & \multicolumn{1}{c}{$P_1$} & \multicolumn{1}{c}{$P_r$} & 
\multicolumn{1}{c}{$P_f$} & \multicolumn{1}{c}{S/N} &  \multicolumn{1}{c}{$P_0$} & \multicolumn{1}{c}{$P_1$} & 
\multicolumn{1}{c}{$P_r$} & \multicolumn{1}{c}{$P_f$} & \multicolumn{1}{c}{S/N} \\
\multicolumn{1}{c}{(1)} & \multicolumn{1}{c}{(2)} & \multicolumn{1}{c}{(3)} &
\multicolumn{1}{c}{(4)} & \multicolumn{1}{c}{(5)} & \multicolumn{1}{c}{(6)} & \multicolumn{1}{c}{(7)} &
\multicolumn{1}{c}{(8)} & \multicolumn{1}{c}{(9)} & \multicolumn{1}{c}{(10)} & \multicolumn{1}{c}{(11)} & \multicolumn{1}{c}{(12)} \\
\hline
NGC\,1407 & 13 3 26 & 0.67 $\pm$ 0.02 &	0.17 & 0.22 & 0.45 &  2.6 &		3.93 $\pm$ 0.07 &	0.14 &  0.29 & 3.64  & 25 \\
NGC\,3258 & 13 6 20 & 0.99 $\pm$ 0.10 &	0.19 & 0.39 & 0.59 &  3.2 &		5.61 $\pm$ 0.20 &	0.12 &  0.27 & 5.34  & 43 \\
NGC\,3268 & 12 6 20 & 0.85 $\pm$ 0.11 &	0.16 & 0.40 & 0.46 &  2.9 &		5.72 $\pm$ 0.08 &	0.11 &  0.25 & 5.46  & 49 \\
NGC\,4696 & 14 6 20 & 0.84 $\pm$ 0.13 &	0.11 & 0.42 & 0.43 &  4.0 &		5.30 $\pm$ 0.19 &	0.07 &  0.26 & 5.04  & 73 \\
NGC\,5322 & 15 6 26 & 0.94 $\pm$ 0.05 &	0.21 & 0.10 & 0.84 &  4.0 &		4.43 $\pm$ 0.06 &	0.18 &  0.18 & 4.26  & 23 \\
NGC\,5557 & 12 3 20 & 0.93 $\pm$ 0.04 &	0.14 & 0.12 & 0.81 &  5.7 &		5.37 $\pm$ 0.20 &	0.13 &  0.18 & 5.19  & 40 \\
\enddata
\label{tab_pi}
\tablenotetext{Notes}{Col. (1): Galaxy; Col. (2) average
annular radius, innermost mask radius, outermost mask radius;
Col. (3-6) $P_0$, $P_1$, $P_r$ and $P_f$ estimated for B-band images;
Col. (7) S/N evaluated as $(P_0-P_r)/P_1$ for B-band frames;
Col. (8-11)  $P_0$, $P_1$, $P_r$ and $P_f$ estimated for I-band images;
Col. (12) S/N  for I-band frames.}
\end{deluxetable}

\begin{deluxetable}{cccccc}
\tabletypesize{\scriptsize}
\rotate
\tablecaption{Observational data}
\tablewidth{0pt}
\startdata
\hline
\multicolumn{1}{c}{Galaxy} & \multicolumn{1}{c}{$\mu_{0,Group}$} & \multicolumn{1}{c}{$B_t$} &  
\multicolumn{1}{c}{$(B{-}I)_0$} & \multicolumn{1}{c}{$\bar{m}_{B,0}$ } &  \multicolumn{1}{c}{$\bar{m}_{I,0}$}  \\
\multicolumn{1}{c}{(1)} & \multicolumn{1}{c}{(2)} & \multicolumn{1}{c}{(3)} &
\multicolumn{1}{c}{(4)} & \multicolumn{1}{c}{(5)} & \multicolumn{1}{c}{(6)}  \\
\hline
NGC\,1407 & 32.01 $\pm$ 0.06 & 10.71 $\pm$ 0.18 & 2.237 $\pm$  0.033  & 34.3 $\pm$ 0.2 & 31.08 $\pm$ 0.07  \\
NGC\,3258 & 32.85 $\pm$ 0.11 & 12.52 $\pm$ 0.13 & 2.189 $\pm$  0.032  & 35.3 $\pm$ 0.3 & 31.95 $\pm$ 0.06  \\
NGC\,3268 & 32.85 $\pm$ 0.11 & 12.30 $\pm$ 0.44 & 2.189 $\pm$  0.032  & 35.5 $\pm$ 0.3 & 31.90 $\pm$ 0.06  \\
NGC\,4696 & 32.95 $\pm$ 0.05 & 11.62 $\pm$ 0.30 & 2.239 $\pm$  0.033  & 35.5 $\pm$ 0.4 & 31.99 $\pm$ 0.07  \\
NGC\,5322 & 32.48 $\pm$ 0.11 & 11.04 $\pm$ 0.22 & 2.105 $\pm$  0.031  & 34.7 $\pm$ 0.1 & 31.22 $\pm$ 0.07  \\
NGC\,5557 & 33.45 $\pm$ 0.11 & 11.92 $\pm$ 0.10 & 2.145 $\pm$  0.032  & 35.2 $\pm$ 0.1 & 32.17 $\pm$ 0.08  \\
\hline
\multicolumn{6}{c}{Data taken from literature} \\
\multicolumn{1}{c}{Galaxy} & \multicolumn{1}{c}{$\mu_{0,Group}$} & \multicolumn{1}{c}{$B_t$} &
\multicolumn{1}{c}{$(V{-}I)_0$} & \multicolumn{1}{c}{$\bar{m}_{B,0}$ } &  \multicolumn{1}{c}{$\bar{m}_{I,0}$}  \\
M\,32      & 24.45 $\pm$ 0.04 & 8.87 $\pm$ 0.35  & 1.133 $\pm$ 0.007 & 26.78 $\pm$ 0.03 & 22.78 $\pm$ 0.04 \\
NGC\,5128 & 27.67 $\pm$ 0.12 & 7.96 $\pm$ 0.26  & 1.078 $\pm$ 0.016 & 30.1  $\pm$ 0.5  & 26.05 $\pm$ 0.11 \\
\enddata
\label{tab_bi}
\tablenotetext{Notes}{Col. (1): Galaxy name; Col. (2) Distance modulus
derived averaging the group $\mu_{0}$ estimations of FP and IRAS
velocity maps distribution (Table 6 data in C05). For NGC\,5557 no
group distance is known, we adopt the weighted average distance of
various $\mu_{0}$ estimated for the galaxy itself. M\,32 distance is
derived as weighted averages of group distances from the
\citet{ferrarese00} database, excluding SBF based distances. NGC\,5128
distance comes from \citet{ferrarese07}, based on Cepheids
variables. Col. (3) : apparent total B magnitude and uncertainty from
Hyperleda Catalogue (URL: http://leda.univ-lyon1.fr); Col. (4):
$(B-I)_0$ integrated color; Col. (5)-(6): B- and I-band SBF apparent
magnitudes.}
\end{deluxetable}

\clearpage

%%%%%%%%%%%%%%%%%%%%%%%%%%%%%%%%%%%%%%%%%%%%%%%%%%%%%%%
%%%%%%%%%%%FIGURES %%%%%%%%%%%%%%%%%%%%%%%%%%%%%%%%%%%%
%%%%%%%%%%%%%%%%%%%%%%%%%%%%%%%%%%%%%%%%%%%%%%%%%%%%%%%

\end{document}